\documentclass[runningheads]{llncs}
\usepackage[utf8]{inputenc}
\usepackage[english]{babel}
\let\proof\relax
 
\usepackage{amsthm}
\usepackage{mathtools}
\usepackage{float}
\usepackage[dvipsnames]{xcolor}

\usepackage{tikz}
\tikzstyle{vertex}=[circle, draw, inner sep=0pt, minimum size=5pt]
\newcommand{\vertex}{\node[vertex]}
\usetikzlibrary{decorations.markings}
\usetikzlibrary{decorations.pathreplacing}
\usetikzlibrary{arrows.meta}
\usepackage{anyfontsize}
\usepackage{graphicx}
\usepackage{amsmath}
\usepackage{amsfonts}
\usepackage{xcolor}
\usepackage{amssymb}
\usepackage{amsmath}
\usepackage{graphicx}
\usepackage{amsmath}
\usepackage{amsfonts}
\usepackage{xcolor}
\newcommand{\boundellipse}[3]
{(#1) ellipse (#2 and #3)
}
\usetikzlibrary{shapes.geometric}
\tikzstyle{square}=[draw, shape=regular polygon, regular polygon sides=4,draw,inner sep=0pt,minimum
size=0.225cm]
\tikzstyle{triangle}=[draw, shape=regular polygon, regular polygon sides=3,draw,inner sep=0pt,minimum
size=0.3cm]

\definecolor{azure}{rgb}{0.0, 0.5, 1.0}
\definecolor{pink}{rgb}{0.84, 0.09, 0.41}
\definecolor{magenta}{rgb}{0.8, 0.0, 0.8}
\definecolor{cyan}{rgb}{0.0, 1.0, 1.0}
\definecolor{green1}{rgb}{0, 1, 0}
\definecolor{green}{rgb}{0, 1, 0}
\definecolor{brown}{rgb}{0.65, 0.16, 0.16}
\definecolor{aquamarine}{rgb}{0.5, 1.0, 0.83}
\definecolor{battleshipgrey}{rgb}{0.52, 0.52, 0.51}
\begin{document}
\title{Defensive  Alliances in Graphs}
%
%
\author{Ajinkya Gaikwad\thanks{The first author  gratefully acknowledges support from the Ministry of Human Resource Development, 
 Government of India, under Prime Minister's Research Fellowship Scheme (No. MRF-192002-211). } \and Soumen Maity\thanks{The 
 second author's research was supported in part by the Science and Engineering Research Board (SERB), Govt. of India, under Sanction Order No.
MTR/2018/00125.}}
\authorrunning{A.\,Gaikwad and S.\,Maity}
%
\institute{Indian Institute of Science Education and Research\\
Dr. Homi Bhabha Road, Pune 411008, INDIA 
\email{\texttt{ajinkya.gaikwad@students.iiserpune.ac.in;}}
\email{\texttt{soumen@iiserpune.ac.in}}
}
\maketitle              
\begin{abstract}  A set $S$ of vertices of a graph is a defensive alliance if, for each element of $S$, the majority of its neighbours are in $S$.   
We study the parameterzied complexity of the 
{\sc Defensive Alliance} problem, where the aim is to find a minimum size
defensive alliance.  Our main results are the following: (1) The {\sc Defensive Alliance} problem has been studied extensively during the last twenty years, but the question whether it is FPT when parameterized by feedback vertex set has still remained open. We prove that the problem is W[1]-hard parameterized by a wide range of fairly restrictive structural parameters such as the feedback vertex set number, treewidth, pathwidth, and treedepth of the input graph;
(2) the problem  parameterized by the vertex cover number of the input graph does not admit a polynomial compression unless coNP $\subseteq$ NP/poly, (3) it
 does not admit $2^{o(n)}$ algorithm under ETH, and   (4) the {\sc Defensive Alliance} problem on circle graphs is NP-complete. 

\end{abstract}

\section{Introduction}
In real life, an alliance is a collection of people, groups, or states such that the union is 
stronger than individual. The alliance can be either to achieve some common purpose, to protect against 
attack, or to assert collective will against others. This motivates the definitions of defensive and offensive
alliances in graphs. 
The properties of alliances in graphs were first studied by Kristiansen, Hedetniemi, and Hedetniemi 
\cite{kris}. 
They introduced defensive, offensive and powerful alliances. An alliance is global  if it is a dominating set. 
The alliance problems have been studied extensively during last fifteen years \cite{frick,SIGARRETA20061345,chel,ROD,SIGA}, and generalizations called $r$-alliances are also studied \cite{SIGARRETA20091687}. 
Throughout this article, $G=(V,E)$ denotes a finite, simple and undirected graph of order $|V|=n$. The subgraph induced by 
$S\subseteq V(G)$ is denoted by $G[S]$. For a vertex $v\in V$, we use $N_G(v)=\{u~:~(u,v)\in E(G)\}$ to denote the (open) neighbourhood 
of vertex $v$ in $G$, and $N_G[v]=N_G(v)\cup \{v\}$ to denote the closed neighbourhood of $v$. The degree $d_G(v)$ of a vertex 
$v\in V(G)$ is $|N_G(v)|$. For a subset $S\subseteq V(G)$, we define its closed neighbourhood as $N_G[S]=\bigcup_{v\in S} N_G[v]$ and its 
open neighbourhood as $N_G(S)=N_G[S]\setminus S$. 
For a non-empty subset $S\subseteq V$ and a vertex $v\in V(G)$, $N_S(v)$ denotes the set of neighbours of $v$ in $S$, that is, 
$N_S(v)=\{ u\in S~:~ (u,v)\in E(G)\}$.  We use $d_S(v)=|N_S(v)|$ to denote the degree of vertex $v$ in $G[S]$. 
The complement of the vertex set $S$ in $V$ is denoted by $S^c$.

\begin{definition}\rm
A non-empty set 
$S\subseteq V$ is a  defensive alliance in $G=(V,E)$ if 
$d_S(v)+1\geq d_{S^c}(v)$ for all $v\in S$.
\end{definition}
\noindent A vertex $v\in S$ is said to be  protected if $d_S(v)+1\geq d_{S^c}(v)$. A set $S\subseteq V$ is a  defensive alliance if every vertex
in $S$ is  protected. 
In this paper, we consider {\sc Defensive Alliance} 
  under structural parameters. We define the problem as follows:
  \vspace{3mm}
    \\
    \fbox
    {\begin{minipage}{33.7em}\label{FFVS }
       {\sc  Defensive Alliance}\\
        \noindent{\bf Input:} An undirected graph $G=(V,E)$ and an  integer $k\geq 1$.
    
        \noindent{\bf Question:} Is there a  defensive alliance $S\subseteq V(G)$ such that $|S|\leq k $?
    \end{minipage} }
    \vspace{3mm}
    \\
    
    
   \noindent  For standard notations and definitions in graph theory and parameterized complexity, we refer to West \cite{west} and Cygan et al. \cite{marekcygan}, respectively. The graph
parameters we explicitly use in this paper are  feedback vertex set number, pathwidth, treewidth and treedepth.  

    \begin{definition} {\rm
        For a graph $G = (V,E)$, the parameter {\it feedback vertex set} is the cardinality of the smallest set $S \subseteq V(G)$ such that the graph $G-S$ is a forest and it is denoted by $fvs(G)$.}
    \end{definition}

\noindent We now review the concept of a tree decomposition, introduced by Robertson and Seymour in \cite{Neil}.
Treewidth is a  measure of how “tree-like” the graph is.
\begin{definition}\rm \cite{Downey} A {\it tree decomposition} of a graph $G=(V,E)$  is a tree $T$ together with a 
collection of subsets $X_t$ (called bags) of $V$ labeled by the vertices $t$ of $T$ such that 
$\bigcup_{t\in T}X_t=V $ and (1) and (2) below hold:
\begin{enumerate}
			\item For every edge $(u,v) \in E(G)$, there  is some $t$ such that $\{u,v\}\subseteq X_t$.
			\item  (Interpolation Property) If $t$ is a vertex on the unique path in $T$ from $t_1$ to $t_2$, then 
			$X_{t_1}\cap X_{t_2}\subseteq X_t$.
		\end{enumerate}
	\end{definition}
	
	
\begin{definition}\rm \cite{Downey} The {\it width} of a tree decomposition is
the maximum value of $|X_t|-1 $ taken over all the vertices $t$ of the tree $T$ of the decomposition.
The treewidth $tw(G)$ of a graph $G$  is the  minimum width among all possible tree decomposition of $G$.
\end{definition} 

\begin{definition}\rm 
    If the tree $T$ of a tree decomposition is a path, then we say that the tree decomposition 
    is a {\it path decomposition}, and use {\it  pathwidth} in place of treewidth. 
\end{definition}

A rooted forest is a disjoint union of rooted trees. Given a rooted forest $F$, its \emph{transitive closure} is a graph $H$ in which $V(H)$ contains all the nodes of the rooted forest, and $E(H)$ contain an edge between two vertices only if those two vertices form an ancestor-descendant pair in the forest $F$.

   \begin{definition}
        {\rm  The {\it treedepth} of a graph $G$ is the minimum height of a rooted forest $F$ whose transitive closure contains the graph $G$. It is denoted by $td(G)$.}
    \end{definition}

 \subsection{Our Main Results}  Our main results are as follows:     
\begin{itemize}
\item the {\sc Defensive Alliance} is W[1]-hard when parameterized by the  
the vertex deletion set into trees of height at most two, even when restricted to
bipartite graphs.
\item the {\sc  Defensive Alliance} problem  parameterized by the vertex cover number of the input graph does not admit a 
 polynomial compression unless coNP $\subseteq$ NP/poly.
 \item the {\sc Defensive Alliance} problem does not admit $2^{o(n)}$ algorithm under ETH.
 \item the {\sc Defensive Alliance} problem on circle graphs is NP-complete.

\end{itemize}    

\subsection{Known Results} The decision version for several types of alliances have been shown to be NP-complete. 
For an integer $r$, a nonempty set $S\subseteq V(G)$ is a {\it defensive $r$-alliance} if for each 
$v\in S$, $|N(v)\cap S|\geq |N(v)\setminus S|+r$. A set is a defensive alliance if it is a defensive 
$(-1)$-alliance. A defensive $r$-alliance $S$ is {\it global} if $S$ is a dominating set. 
 The defensive $r$-alliance problem   is NP-complete for any $r$ \cite{SIGARRETA20091687}. The defensive alliance problem is 
 NP-complete even when restricted to split, chordal and bipartite graph \cite{Lindsay}. 
 For an integer $r$, a nonempty set $S\subseteq V(G)$ is an {\it offensive $r$-alliance} if for each 
$v\in N(S)$, $|N(v)\cap S|\geq |N(v)\setminus S|+r$. An offensive 1-alliance is called an offensive
alliance.  An offensive $r$-alliance $S$ is {\it global} if $S$ is a dominating set. 
 Fernau et al. showed that the offensive $r$-alliance and global 
 offensive $r$-alliance problems are NP-complete for any fixed $r$ \cite{FERNAU2009177}. 
 They also proved that for $r>1$, $r$-offensive alliance is NP-hard, even when restricted to 
 $r$-regular planar graphs.  There are polynomial time algorithms for finding minimum alliances
 in trees \cite{CHANG2012479,Lindsay}.  A polynomial time algorithm for finding minimum defensive alliance in series parallel graph is presented in \cite{10.5555/1292785}. Fernau  and Raible showed in \cite{Fernau} that the defensive, offensive and 
 powerful alliance problems and their global
 variants are fixed parameter tractable when parameterized by solution size $k$. Kiyomi and Otachi
 showed in 
 \cite{KIYOMI201791}, the problems of finding smallest alliances of all kinds are fixed-parameter tractable
 when parameteried by the vertex cover number. The problems of finding smallest defensive 
 and offensive alliances are also fixed-parameter tractable
 when parameteried by the neighbourhood diversity \cite{ICDCIT2021}. 
 Enciso \cite{Enciso2009AlliancesIG} proved that 
 finding defensive and global defensive alliances is fixed parameter tractable when parameterized by domino treewidth. 
 Bliem and Woltran \cite{BLIEM2018334} proved that 
  deciding if a graph contains a defensive alliance of size at most
$k$
 is W[1]-hard when parameterized by treewidth of the input graph. This puts it among the few problems that are FPT when parameterized by solution size but not when parameterized 
by treewidth (unless FPT=W[1]).

\section{Hardness Results of Defensive Alliance}\label{hardnesssection}
In this section  we show that  the {\sc Defensive Alliance} problem is W[1]-hard when parameterized by  
the size of a vertex deletion set into trees of height at most 2, even when restricted to bipartite graphs, via a reduction 
from  the {\sc Multidimensional Relaxed Subset Sum (MRSS)} problem. 
\noindent\vspace{3mm}
    \\
    \fbox
    {\begin{minipage}{33.7em}\label{SP1}
       {\sc  Multidimensional Subset Sum (MSS)}\\
     \noindent{\bf  Input:} An integer $k$, a set 
     $S = \{s_1,\ldots,s_n\}$ of vectors with $s_i \in \mathbb{N}^k$ for every $i$ with 
     $1 \leq i \leq  n$  and a target vector $t \in \mathbb{N}^k$.\\
\noindent {\bf Parameter}: $k$ \\
\noindent{\bf Question}: Is there a subset $S'\subseteq S $ such that $\sum\limits_{s\in S'}{s}=t$?
    \end{minipage} }\\

  \noindent  We consider a variant of MSS that we require in our proofs. 
  In the {\sc Multidimensional Relaxed Subset Sum} ({\sc MRSS}) problem, an additional integer $k'$ is given
    (which will be part of the parameter)
    and we ask whether there is a subset $S'\subseteq S$ with $|S'|\leq k'$ such that $\sum\limits_{s\in S'}{s}\geq t$.
    This variant can be formalized as
follows:

\noindent\vspace{3mm}
    \\
    \fbox
    {\begin{minipage}{33.7em}\label{SP3}
       {\sc Multidimensional Relaxed Subset Sum (MRSS)}\\
     \noindent{\bf  Input:} An integer $k$, a set 
     $S = \{s_1,\ldots,s_n\}$ of vectors with $s_i \in \mathbb{N}^k$ for every $i$ with 
     $1 \leq i \leq  n$, a target vector $t \in \mathbb{N}^k$ and an integer $k'$.\\
\noindent {\bf Parameter}: $k+k'$ \\
\noindent{\bf Question}: Is there a subset $S'\subseteq S $ with $|S'|\leq k'$ such that $\sum\limits_{s\in S'}{s}\geq t$?
    \end{minipage} }\\
    
\noindent It is known that {\sc MRSS} is W[1]-hard when parameterized by the combined parameter $k+k'$,
   even if all integers in the input are given in unary \cite{mss}.
\noindent We now show that the {\sc Defensive Alliance} problem is W[1]-hard parameterized by  the size of a vertex 
 deletion set into trees of height at most 2, via a reduction from  MRSS. We now prove the following theorem:
 \begin{theorem}\label{FNvds}\rm
 The {\sc Defensive Alliance} problem is W[1]-hard when parameterized by  the size of a vertex deletion set into 
 trees of height at most 2, even when restricted to bipartite graphs.
 \end{theorem}
 
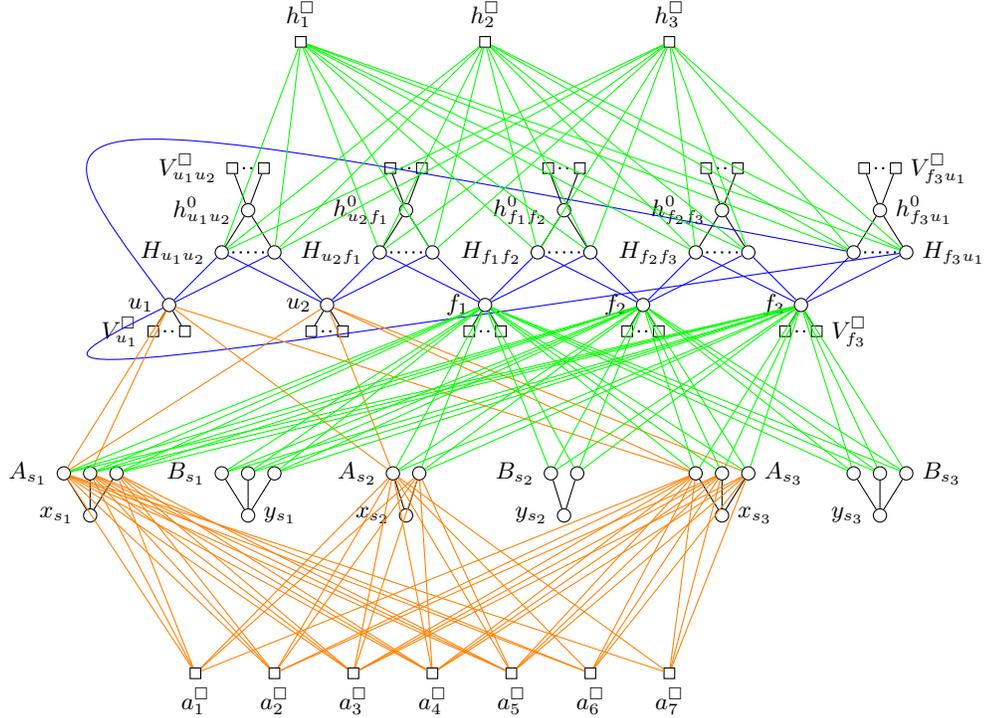
\begin{figure}[ht]
    \centering
   \begin{tikzpicture}[scale=0.7]
\node[circle, draw, inner sep=0pt, minimum size=5pt] (u1) at (0,4) [label=left:$u_1$] {};
\node[circle, draw, inner sep=0pt, minimum size=5pt] (u2) at (3,4) [label=left:$u_2$] {};
\node[circle, draw, inner sep=0pt, minimum size=5pt] (f1) at (6,4) [label=left:$f_1$] {};
\node[circle, draw, inner sep=0pt, minimum size=5pt] (f2) at (9,4) [label=left:$f_2$] {};
\node[circle, draw, inner sep=0pt, minimum size=5pt] (f3) at (12,4) [label=left:$f_3$] {};

\node[draw,square, inner sep=0pt, minimum size=0.2cm] (u11) at (-0.3,3.5) [label=left: $V_{u_1}^{\square}$] {};
\node[draw,square, inner sep=0pt, minimum size=0.2cm] (u12) at (0.3,3.5) [label=below:] {};
\draw [thick, dotted] (-0.1,3.5) -- (0.1,3.5);

\node[draw,square, inner sep=0pt, minimum size=0.2cm] (u21) at (2.7,3.5) [label=below:] {};
\node[draw,square, inner sep=0pt, minimum size=0.2cm] (u22) at (3.3,3.5) [label=below:] {};
\draw [thick, dotted] (2.9,3.5) -- (3.1,3.5);

\node[draw,square, inner sep=0pt, minimum size=0.2cm] (f11) at (5.7,3.5) [label=below:] {};
\node[draw,square, inner sep=0pt, minimum size=0.2cm] (f12) at (6.3,3.5) [label=below:] {};
\draw [thick, dotted] (5.9,3.5) -- (6.1,3.5);

\node[draw,square, inner sep=0pt, minimum size=0.2cm] (f21) at (8.7,3.5) [label=below:] {};
\node[draw,square, inner sep=0pt, minimum size=0.2cm] (f22) at (9.3,3.5) [label=below:] {};
\draw [thick, dotted] (8.9,3.5) -- (9.1,3.5);

\node[draw,square, inner sep=0pt, minimum size=0.2cm] (f31) at (11.7,3.5) [label=below:] {};
\node[draw,square, inner sep=0pt, minimum size=0.2cm] (f32) at (12.3,3.5) [label=right: $V_{f_3}^{\square}$] {};
\draw [thick, dotted] (11.9,3.5) -- (12.1,3.5);

\path
(u1) edge  (u11)
(u1) edge (u12)
(u2) edge (u21)
(u2) edge (u22)
(f1) edge (f11)
(f1) edge (f12)
(f2) edge (f21)
(f2) edge (f22)
(f3) edge (f31)
(f3) edge (f32);
\node[circle, draw, inner sep=0pt, minimum size=5pt] (xs1) at (-1.5,0) [label=left:$x_{s_1}$] {};
\node[circle, draw, inner sep=0pt, minimum size=5pt] (xs11) at (-2,0.8) [label=left:$A_{s_1}$] {};
\node[circle, draw, inner sep=0pt, minimum size=5pt] (xs12) at (-1.5,0.8) [] {};
\node[circle, draw, inner sep=0pt, minimum size=5pt] (xs13) at (-1,0.8) [] {};

\path 
(xs1) edge (xs11)
(xs1) edge (xs12)
(xs1) edge (xs13);

\node[circle, draw, inner sep=0pt, minimum size=5pt] (ys1) at (1.5,0) [label=right:$y_{s_1}$] {};
\node[circle, draw, inner sep=0pt, minimum size=5pt] (ys11) at (1,0.8) [label=left:$B_{s_1}$] {};
\node[circle, draw, inner sep=0pt, minimum size=5pt] (ys12) at (1.5,0.8) [] {};
\node[circle, draw, inner sep=0pt, minimum size=5pt] (ys13) at (2,0.8) [] {};
\path 
(ys1) edge (ys11)
(ys1) edge (ys12)
(ys1) edge (ys13);

\node[circle, draw, inner sep=0pt, minimum size=5pt] (xs2) at (4.5,0) [label=left:$x_{s_2}$] {};
\node[circle, draw, inner sep=0pt, minimum size=5pt] (xs21) at (4.25,0.8) [label=left:$A_{s_2}$] {};
\node[circle, draw, inner sep=0pt, minimum size=5pt] (xs22) at (4.75,0.8) [] {};
\path
(xs2) edge (xs21)
(xs2) edge (xs22);

\node[circle, draw, inner sep=0pt, minimum size=5pt] (ys2) at (7.5,0) [label=left:$y_{s_2}$] {};
\node[circle, draw, inner sep=0pt, minimum size=5pt] (ys21) at (7.25,0.8) [label=left:$B_{s_2}$] {};
\node[circle, draw, inner sep=0pt, minimum size=5pt] (ys22) at (7.75,0.8) [] {};
\path
(ys2) edge (ys21)
(ys2) edge (ys22);

\node[circle, draw, inner sep=0pt, minimum size=5pt] (xs3) at (10.5,0) [label=right:$x_{s_3}$] {};
\node[circle, draw, inner sep=0pt, minimum size=5pt] (xs31) at (10,0.8) [] {};
\node[circle, draw, inner sep=0pt, minimum size=5pt] (xs32) at (10.5,0.8) [] {};
\node[circle, draw, inner sep=0pt, minimum size=5pt] (xs33) at (11,0.8) [label=right:$A_{s_3}$] {};
\path 
(xs3) edge (xs31)
(xs3) edge (xs32)
(xs3) edge (xs33);

\node[circle, draw, inner sep=0pt, minimum size=5pt] (ys3) at (13.5,0) [label=left:$y_{s_3}$] {};
\node[circle, draw, inner sep=0pt, minimum size=5pt] (ys31) at (13,0.8) [] {};
\node[circle, draw, inner sep=0pt, minimum size=5pt] (ys32) at (13.5,0.8) [] {};
\node[circle, draw, inner sep=0pt, minimum size=5pt] (ys33) at (14,0.8) [label=right:$B_{s_3}$] {};
\path 
(ys3) edge (ys31)
(ys3) edge (ys32)
(ys3) edge (ys33);

\node[circle, draw, inner sep=0pt, minimum size=5pt] (u1u21) at (1,5) [label=left: $H_{u_1u_2}$] {};
\node[circle, draw, inner sep=0pt, minimum size=5pt] (u1u22) at (2,5) [] {};
\draw [thick, dotted] (1.2,5) -- (1.8,5);

\node[circle, draw, inner sep=0pt, minimum size=5pt] (h0u1u2) at (1.5,5.8) [label=left:$h^0_{u_1u_2}$] {};
\node[draw,square, inner sep=0pt, minimum size=0.2cm] (h1u1u2) at (1.2,6.6) [label=left: $V_{u_1u_2}^{\square}$] {};
\node[draw,square, inner sep=0pt, minimum size=0.2cm] (h2u1u2) at (1.8,6.6) [label=below:] {};
\draw [thick, dotted] (1.4,6.6) -- (1.6,6.6);

\path
(h0u1u2) edge (h1u1u2)
(h0u1u2) edge (h2u1u2)
(h0u1u2) edge (u1u21)
(h0u1u2) edge (u1u22)
(u1) edge [color=blue]  (u1u21)
(u1) edge [color=blue] (u1u22)
(u2) edge [color=blue] (u1u21)
(u2) edge [color=blue] (u1u22);

\node[circle, draw, inner sep=0pt, minimum size=5pt] (u2f11) at (4,5) [label=left:$H_{u_2f_1}$] {};
\node[circle, draw, inner sep=0pt, minimum size=5pt] (u2f12) at (5,5) [] {};
\draw [thick, dotted] (4.2,5) -- (4.8,5);

\node[circle, draw, inner sep=0pt, minimum size=5pt] (h0u2f1) at (4.5,5.8) [label=left:$h^0_{u_2f_1}$] {};
\node[draw,square, inner sep=0pt, minimum size=0.2cm] (h1u2f1) at (4.2,6.6) [label=left: ] {};
\node[draw,square, inner sep=0pt, minimum size=0.2cm] (h2u2f1) at (4.8,6.6) [label=below:] {};
\draw [thick, dotted] (4.4,6.6) -- (4.6,6.6);

\path
(h0u2f1) edge (h1u2f1)
(h0u2f1) edge (h2u2f1)
(h0u2f1) edge (u2f11)
(h0u2f1) edge (u2f12)
(u2) edge [color=blue] (u2f11)
(u2) edge [color=blue] (u2f12)
(f1) edge [color=blue] (u2f11)
(f1) edge [color=blue] (u2f12);

\node[circle, draw, inner sep=0pt, minimum size=5pt] (f1f21) at (7,5) [label=left:$H_{f_1f_2}$] {};
\node[circle, draw, inner sep=0pt, minimum size=5pt] (f1f22) at (8,5) [] {};
\draw [thick, dotted] (7.2,5) -- (7.8,5);

\node[circle, draw, inner sep=0pt, minimum size=5pt] (h0f1f2) at (7.5,5.8) [label=left:$h^0_{f_1f_2}$] {};
\node[draw,square, inner sep=0pt, minimum size=0.2cm] (h1f1f2) at (7.2,6.6) [label=below:] {};
\node[draw,square, inner sep=0pt, minimum size=0.2cm] (h2f1f2) at (7.8,6.6) [label=below:] {};
\draw [thick, dotted] (7.4,6.6) -- (7.6,6.6);

\path
(h0f1f2) edge (h1f1f2)
(h0f1f2) edge (h2f1f2)
(h0f1f2) edge (f1f21)
(h0f1f2) edge (f1f22)
(f1) edge [color=blue] (f1f21)
(f1) edge [color=blue] (f1f22)
(f2) edge [color=blue] (f1f21)
(f2) edge [color=blue] (f1f22);

\node[circle, draw, inner sep=0pt, minimum size=5pt] (f2f31) at (10,5) [label=left:$H_{f_2f_3}$] {};
\node[circle, draw, inner sep=0pt, minimum size=5pt] (f2f32) at (11,5) [] {};
\draw [thick, dotted] (10.2,5) -- (10.8,5);

\node[circle, draw, inner sep=0pt, minimum size=5pt] (h0f2f3) at (10.5,5.8) [label=left:$h^0_{f_2f_3}$] {};
\node[draw,square, inner sep=0pt, minimum size=0.2cm] (h1f2f3) at (10.2,6.6) [label=below:] {};
\node[draw,square, inner sep=0pt, minimum size=0.2cm] (h2f2f3) at (10.8,6.6) [label=below:] {};
\draw [thick, dotted] (10.4,6.6) -- (10.6,6.6);

\path
(h0f2f3) edge (h1f2f3)
(h0f2f3) edge (h2f2f3)
(h0f2f3) edge (f2f31)
(h0f2f3) edge (f2f32)
(f2) edge [color=blue] (f2f31)
(f2) edge [color=blue](f2f32)
(f3) edge [color=blue] (f2f31)
(f3) edge [color=blue] (f2f32);

\node[circle, draw, inner sep=0pt, minimum size=5pt] (f3u11) at (13,5) [] {};
\node[circle, draw, inner sep=0pt, minimum size=5pt] (f3u12) at (14,5) [label=right:$H_{f_3u_1}$] {};
\draw [thick, dotted] (13.2,5) -- (13.8,5);

\node[circle, draw, inner sep=0pt, minimum size=5pt] (h0f3u1) at (13.5,5.8) [label=right:$h^0_{f_3u_1}$] {};
\node[draw,square, inner sep=0pt, minimum size=0.2cm] (h1f3u1) at (13.2,6.6) [label=below:] {};
\node[draw,square, inner sep=0pt, minimum size=0.2cm] (h2f3u1) at (13.8,6.6) [label=right:$V_{f_3u_1}^{\square}$] {};
\draw [thick, dotted] (13.4,6.6) -- (13.6,6.6);

\path
(h0f3u1) edge (h1f3u1)
(h0f3u1) edge (h2f3u1)
(h0f3u1) edge (f3u11)
(h0f3u1) edge (f3u12)
(f3) edge [color=blue] (f3u11)
(f3) edge [color=blue] (f3u12);

\draw[color=blue] (u1) .. controls (-3,8) ..  (f3u11);
\draw[color=blue] (u1) ..controls (-3,2.5).. (f3u12);

\node[draw,square, inner sep=0pt, minimum size=0.2cm] (h1s) at (2.5,9) [label=above: $h_1^{\square}$] {};
\node[draw,square, inner sep=0pt, minimum size=0.2cm] (h2s) at (6,9) [label=above: $h_2^{\square}$] {};
\node[draw,square, inner sep=0pt, minimum size=0.2cm] (h3s) at (9.5,9) [label=above: $h_3^{\square}$] {};

\path
(h1s) edge [color=green] (u1u21)
(h1s) edge [color=green] (u1u22)
(h1s) edge [color=green] (u2f11)
(h1s) edge [color=green] (u2f12)
(h1s) edge [color=green] (f1f21)
(h1s) edge [color=green] (f1f22)
(h1s) edge [color=green] (f2f31)
(h1s) edge [color=green] (f2f32)
(h1s) edge [color=green] (f3u11)
(h1s) edge [color=green] (f3u12)
(h2s) edge [color=green] (u1u21)
(h2s) edge [color=green] (u1u22)
(h2s) edge [color=green] (u2f11)
(h2s) edge [color=green] (u2f12)
(h2s) edge [color=green] (f1f21)
(h2s) edge [color=green] (f1f22)
(h2s) edge [color=green] (f2f31)
(h2s) edge [color=green] (f2f32)
(h2s) edge [color=green] (f3u11)
(h2s) edge [color=green] (f3u12)
(h3s) edge [color=green] (u1u21)
(h3s) edge [color=green] (u1u22)
(h3s) edge [color=green] (u2f11)
(h3s) edge [color=green] (u2f12)
(h3s) edge [color=green] (f1f21)
(h3s) edge [color=green] (f1f22)
(h3s) edge [color=green] (f2f31)
(h3s) edge [color=green] (f2f32)
(h3s) edge [color=green] (f3u11)
(h3s) edge [color=green] (f3u12);

\path
(u1) edge [color=orange] (xs11)
(u1) edge [color=orange] (xs12)
(u1) edge [color=orange] (xs21)
(u1) edge [color=orange] (xs31)
(u2) edge [color=orange] (xs11)
(u2) edge [color=orange] (xs21)
(u2) edge [color=orange] (xs31)
(u2) edge [color=orange] (xs33)
;

\path 
(f1) edge [color=green] (xs11)
(f1) edge [color=green] (xs12)
(f1) edge [color=green] (xs13)
(f1) edge [color=green] (ys11)
(f1) edge [color=green] (ys12)
(f1) edge [color=green] (ys13)
(f1) edge [color=green] (xs21)
(f1) edge [color=green] (xs22)
(f1) edge [color=green] (ys21)
(f1) edge [color=green] (ys22)
(f1) edge [color=green] (xs31)
(f1) edge [color=green] (xs32)
(f1) edge [color=green] (xs33)
(f1) edge [color=green] (ys31)
(f1) edge [color=green] (ys32)
(f1) edge [color=green] (ys33)
(f2) edge [color=green] (xs11)
(f2) edge [color=green] (xs12)
(f2) edge [color=green] (xs13)
(f2) edge [color=green] (ys11)
(f2) edge [color=green] (ys12)
(f2) edge [color=green] (ys13)
(f2) edge [color=green] (xs21)
(f2) edge [color=green] (xs22)
(f2) edge [color=green] (ys21)
(f2) edge [color=green] (ys22)
(f2) edge [color=green] (xs31)
(f2) edge [color=green] (xs32)
(f2) edge [color=green] (xs33)
(f2) edge [color=green] (ys31)
(f2) edge [color=green] (ys32)
(f2) edge [color=green] (ys33)
(f3) edge [color=green] (xs11)
(f3) edge [color=green] (xs12)
(f3) edge [color=green] (xs13)
(f3) edge [color=green] (ys11)
(f3) edge [color=green] (ys12)
(f3) edge [color=green] (ys13)
(f3) edge [color=green] (xs21)
(f3) edge [color=green] (xs22)
(f3) edge [color=green] (ys21)
(f3) edge [color=green] (ys22)
(f3) edge [color=green] (xs31)
(f3) edge [color=green] (xs32)
(f3) edge [color=green] (xs33)
(f3) edge [color=green] (ys31)
(f3) edge [color=green] (ys32)
(f3) edge [color=green] (ys33);

\node[draw,square, inner sep=0pt, minimum size=0.2cm] (h1s) at (0.5,-3) [label=below:$a_1^{\square}$] {};
\node[draw,square, inner sep=0pt, minimum size=0.2cm] (h2s) at (2,-3) [label=below:$a_2^{\square}$] {};
\node[draw,square, inner sep=0pt, minimum size=0.2cm] (h3s) at (3.5,-3) [label=below:$a_3^{\square}$] {};
\node[draw,square, inner sep=0pt, minimum size=0.2cm] (h4s) at (5,-3) [label=below:$a_4^{\square}$] {};
\node[draw,square, inner sep=0pt, minimum size=0.2cm] (h5s) at (6.5,-3) [label=below:$a_5^{\square}$] {};
\node[draw,square, inner sep=0pt, minimum size=0.2cm] (h6s) at (8,-3) [label=below:$a_6^{\square}$] {};
\node[draw,square, inner sep=0pt, minimum size=0.2cm] (h7s) at (9.5,-3) [label=below:$a_7^{\square}$] {};
\path
(xs11) edge [color=orange] (h1s)
(xs11) edge [color=orange] (h2s)
(xs11) edge [color=orange] (h3s)
(xs11) edge [color=orange] (h4s)
(xs11) edge [color=orange] (h5s)
(xs11) edge [color=orange] (h6s)
(xs11) edge [color=orange] (h7s)

(xs12) edge [color=orange] (h1s)
(xs12) edge [color=orange] (h2s)
(xs12) edge [color=orange] (h3s)
(xs12) edge [color=orange] (h4s)
(xs12) edge [color=orange] (h5s)
(xs12) edge [color=orange] (h6s)

(xs13) edge [color=orange] (h1s)
(xs13) edge [color=orange] (h2s)
(xs13) edge [color=orange] (h3s)
(xs13) edge [color=orange] (h4s)
(xs13) edge [color=orange] (h5s)

(xs21) edge [color=orange] (h1s)
(xs21) edge [color=orange] (h2s)
(xs21) edge [color=orange] (h3s)
(xs21) edge [color=orange] (h4s)
(xs21) edge [color=orange] (h5s)
(xs21) edge [color=orange] (h6s)
(xs21) edge [color=orange] (h7s)

(xs22) edge [color=orange] (h1s)
(xs22) edge [color=orange] (h2s)
(xs22) edge [color=orange] (h3s)
(xs22) edge [color=orange] (h4s)
(xs22) edge [color=orange] (h5s)

(xs31) edge [color=orange] (h1s)
(xs31) edge [color=orange] (h2s)
(xs31) edge [color=orange] (h3s)
(xs31) edge [color=orange] (h4s)
(xs31) edge [color=orange] (h5s)
(xs31) edge [color=orange] (h6s)
(xs31) edge [color=orange] (h7s)

(xs32) edge [color=orange] (h2s)
(xs32) edge [color=orange] (h3s)
(xs32) edge [color=orange] (h4s)
(xs32) edge [color=orange] (h5s)
(xs32) edge [color=orange] (h6s)
(xs32) edge [color=orange] (h7s)

(xs33) edge [color=orange] (h3s)
(xs33) edge [color=orange] (h4s)
(xs33) edge [color=orange] (h5s)
(xs33) edge [color=orange] (h6s)
(xs33) edge [color=orange] (h7s)

;

\end{tikzpicture}
\caption{The graph $G$ in the proof of Theorem \ref{FNvds} constructed for MRSS instance 
 $S=\{(2, 1), (1,1), (1,2)\}$, $t=(3,3)$, $k=2$ and $k'=2$. The vertices $t$ and $t'$, and their adjacency are 
 not shown. Every square vertex is adjacent to $2r+2$ vertices which are also not shown in the diagram. }
\label{intractibilityDA1}
\end{figure}

\proof 
Let  $I = (k, k', S, t)$  be an instance of {\sc MRSS}.  From this we construct an  instance 
 $I'=(G,r)$ of {\sc Defensive Alliance} the following way. 
 See Figure \ref{intractibilityDA1} for an illustration.
Let $s=(s(1),s(2),\ldots,s(k))\in S$ and let $\text{max}(s)$ denote  the value of the largest coordinate of $s$.
We set $N= \sum\limits_{s\in S } (2\text{max}(s) +2)$.
First we  introduce a  set  of $k$ new vertices $U=\{ u_1,u_2,\ldots,u_k\}$.
  For every $u_i\in U$, we introduce a set $V_{u_i}^{\square}$ of $\sum\limits_{s\in S}{s(i)} + 2N - 2(\sum\limits_{s\in S}{s(i)} - t(i)) $  square vertices and   make $u_i$ adjacent to every vertex 
 of $V_{u_i}^{\square}$.   We also introduce a  set $F =\{f_{1},f_{2},f_{3}\}$ of three new vertices and make each $f\in F$ adjacent to a 
 set $V_{f}^{ \square}$ of $2N$ square vertices.  
 We consider the vertices of $U$ and $F$ in the following order: $u_{1},u_2,\ldots,u_k,f_{1},f_{2},f_{3}, u_1$. 
 Let $$P=\{(u_1,u_2),\ldots,(u_{k-1},u_k),(u_{k},f_{1}),(f_{1},f_{2}),(f_{2},f_{3}),(f_{3},u_{1})\}$$ be the
 set of pairs of consecutive vertices.
 For each  pair $(x,y)\in P$, we  add a  set $H_{xy}=\{h^1_{xy},\ldots,h^N_{xy}\}$ of $N$ new vertices and add a 
 special vertex $h^{0}_{xy}$ which is adjacent to all the vertices of $H_{xy}$; 
 $h^0_{xy}$ is also adjacent to a  set $V_{xy}^{\square}$ of $N$ new square vertices.
 Finally, we introduce a set $H^{\square} = \{h_{1}^{\square},h_{2}^{\square},h_{3}^{\square}\}$ of three new square 
 vertices and for each pair $(x,y)\in P$, make 
 vertices of $H_{xy}$ adjacent to  square vertices of $H^{\square}$.

Recall that  a star $S_k$ is the complete bipartite graph $K_{1,k}$: a tree with one internal node and $k$ leaves.
For each vector
$s\in S$, we introduce two stars  $S^1_{\mbox{max}(s)+1}$ and $S^2_{\mbox{max}(s)+1}$.
The first star has internal node $x_s$ and $\mbox{max}(s)+1$ leaves $A_s=\{a_1^s,\ldots,a_{\text{max}(s)+1}^s \}$;
the second star has internal node $y_s$ and $\mbox{max}(s)+1$ leaves $B_s=\{b_1^s,\ldots,b_{\text{max}(s)+1}^s \}$.
For each $i\in \{1,2,\ldots,k\}$ and 
 for each $s\in S$, we make $u_i$ adjacent to exactly $s(i)$ many vertices 
 of  $A_s$ in an arbitrary manner. We make the three vertices $f_{1},f_{2},f_{3}$ from the set $F$ 
 adjacent to all vertices of  $\bigcup\limits_{s\in S} {A_s\cup B_s}$.

Finally we add a set $V_{a}^{\square} = \{ a_{1}^{\square},\ldots, a_{k+5}^{\square}\}$  of $k+5$ square vertices. 
We make each vertex in $A_{s}$ adjacent to exactly $|N_{U}(a^{s})|+5$  many vertices of $V_{a}^{\square}$
arbitrarily.
We define the set of square vertices as
$V_{\square} = {V_{a}^{\square}} \cup H^{\square} \bigcup\limits_{(x,y)\in P} V_{xy}^{\square} \bigcup\limits_{u\in U} V_{u}^{\square} \bigcup\limits_{f\in F} V_{f}^{\square} $ and  we also define $  V_{\triangle} = U \cup F \bigcup\limits_{(x,y)\in P} H_{xy}\cup \{h^{0}_{xy}\}$. We set 
$r= (k+3)N + 2k+6 +\sum\limits_{i=1}^{n}(\text{max}(s_{i})+1)+k'$.
For every vertex $x\in V_{\square}$, we introduce a set 
 $V_x=\{x_1,x_2,\dots,x_{2r+2}\}$ of $2r+2$ many vertices adjacent to $x$.  
 We also add two vertices $t$ and $t'$. The vertex $t$ is adjacent to all the vertices in $\bigcup \{V_{x} ~|~ x\in \bigcup\limits_{(x,y)\in P} V_{xy}^{\square} \bigcup\limits_{u\in U} V_{u}^{\square} \bigcup\limits_{f\in F} V_{f}^{\square} \}$. Similarly, the vertex $t'$ is adjacent to all the vertices in  $\bigcup \{V_{x} ~|~ x\in {V_{a}^{\square}} \cup H^{\square}\}$. This completes the construction of graph $G$. 
 The strategy is to force all the square vertices outside the solution and all the vertices in $V_{\triangle}$ inside the solution. Observe that if we remove the set $U\cup F \cup H^{\square} \cup V_{a}^{\square} \cup \{ h^0_{xy} ~|~ (x,y)\in P \} \bigcup \{t,t'\}  $ of $3k+16$ vertices from $G$ then we are left with only star graphs. It is easy to see  that $G$ is a 
 bipartite graph with bipartition  $$V_{1} =  H^{\square} \cup V_{a}^{\square} \cup U \cup F\cup \{t'\} \bigcup\limits_{x \in V_{\square}\setminus (H^{\square} \cup V_{a}^{\square})} V_{x} \bigcup\limits_{(x,y)\in P} \{h^0_{xy}\} \bigcup\limits_{s \in S} \{x_{s},y_{s}\}  $$ and $$V_{2} = \{t\} \bigcup\limits_{(x,y)\in P} (V_{xy}^{\square} \cup H_{xy}) \bigcup\limits_{u\in U} V_{u}^{\square} \bigcup\limits_{f\in F} V_{f}^{\square} \bigcup\limits_{x \in H^{\square} \cup V_{a}^{\square}} V_{x} \bigcup\limits_{s\in S} (A_{s} \cup B_{s}).$$
 
\noindent We claim that $I = (k, k', S, t)$ is a positive instance of {\sc MRSS} if and only if $I'=(G,r)$ is a positive instance of {\sc Defensive
Alliance}.
Let $S'\subseteq S$ be such that $|S'|\leq k'$ and $\sum\limits_{s\in S'}{s}\geq t$. We claim that the set
\begin{align*}
    R =& U \cup F \bigcup\limits_{(x,y)\in P} H_{xy}\cup \{h^{0}_{xy}\} \bigcup\limits_{s\in S'} A_s \cup\{x_s\} \bigcup\limits_{s\in S\setminus S'} B_{s} 
\end{align*} 
is a defensive alliance in $G$ such that $|R|\leq r$. 
Let $x$ be an arbitrary element of $R$. \\
{\it Case 1:} If $x=u_i\in U$, then 
\begin{equation*}
 d_R(u_i)=\sum\limits_{s \in S'} s(i) + 2N
\end{equation*}
and 
\begin{equation*}
 d_{R^c}(u_i)=\sum\limits_{s \in S\backslash S'} s(i) +|V_{u_i}^{\square}|
 =\sum\limits_{s \in S\backslash S'} s(i) + \sum\limits_{s \in S} s(i) + 2N - 2(\sum\limits_{s\in S} s(i)-t(i))
\end{equation*}

\begin{equation*}
\begin{split}
 d_{R^{c}}(u_i)&=2N + \sum\limits_{s \in S\backslash S'} s(i) + \sum\limits_{s \in S} s(i) - 2(\sum\limits_{s\in S} s(i)-t(i))\\
           &=2N -\Big( \sum\limits_{s \in S} s(i) - \sum\limits_{s\in S\backslash S'} s(i)\Big) + 2t(i)  \\
           &=2N -\sum\limits_{s \in S'} s(i) + 2t(i)\\
           &= 2N + \sum\limits_{s\in S'} s(i) + 2\Big(t(i)-\sum\limits_{s\in S'}s(i) \Big)  \\
           &\leq 2N + \sum\limits_{s\in S'} s(i) \\
           &= d_{R}(u_i)
 \end{split}          
\end{equation*}
Therefore, we have $d_R(u_i)+1\geq d_{R^c}(u_i)$, and hence $u_i$ is protected. \\

\noindent {\it Case 2:} If  $x=a^{s}\in A_s$, then  $d_{R}(a^s) = |N_{U}(a^s)| + |\{f_{1},f_{2},f_{3},x_s\}|=|N_{U}(a^s)|+4$ and 
$d_{R^c}(a^s) =|N_{U}(a^s)|+5$. Therefore, we get $d_{R}(a^s)+1 \geq d_{R^c}(a^s)$.\\\\
{\it Case 3:} If $x=f_{1} \in F$, then $N_{R}(f_{1})=\bigcup\limits_{s\in S'} A_{s} \bigcup\limits_{s\in S\backslash S'} B_{s} \cup H_{u_2f_1} \cup H_{f_1f_2}$ and 
$N_{R^c} (a)=\bigcup\limits_{s\in S'} B_{s} \bigcup\limits_{s\in S\backslash S'} A_{s} \cup V_{f_{1}}^{\square} $. 
As $|A_s|=|B_s|$, $|H_{u_2f_1}|=|H_{f_1f_2}|=N$ and $|V_{f_1}^{\square}|=2N+1$, we have 
$d_{R}(f_1)+1\geq d_{R^c}(f_1)$. We can similarly check that $\{f_{2},f_{2}\}$ are also protected.\\

\noindent For the rest of the vertices in $R$, it is easy to see that $d_R(x)+1\geq d_{R^c}(x)$. 
Therefore, $I'=(G,r)$ is a 
yes instance.

\par For the reverse direction, suppose that $G$ has a defensive alliance 
$R$ of size at most $r$. 
It is easy to see that $(V_{\square} \cup \{t,t'\}) \cap R=\emptyset$ as any defensive 
alliance of size at most $r$ cannot contain vertices of degree greater than $2r$. 
This also shows that $ \bigcup\limits_{x\in V_{\square}} {V_{x} \cap R =\emptyset}$.
Now we  show that $  V_{\triangle} = U \cup F \bigcup\limits_{(x,y)\in P} H_{xy}\cup \{h^{0}_{xy}\} \subseteq R$. 
We claim that if $V_{\triangle}\cap R \neq \emptyset$ then $V_{\triangle}\subseteq R$. \\\\
{\it Case 1:} Suppose $R$ contains $u_1$ from $U$. We observe that if 
some $h_{u_1u_2}\in H_{u_1u_2}$  is not in $R$ then $h^{0}_{u_1u_2}$ is also not in 
$R$. 
This implies that no vertex from  $H_{u_1u_2}$ is in $R$. 
In this case $N_{R^{c}}(u_{1})\geq 3N + 2t(i) -\sum\limits_{s\in S} s(i)$ and $N_{R}(u_{1}) \leq N + \sum\limits_{i=1}^{n}(\text{max}(s_{i})+2)$. 
This implies that $u_{1}$ is not protected in $R$ which is a contradiction as $u_{1}\in R$. 
This implies that $H_{u_1u_2}\cup \{h^{0}_{u_1u_2}\} \subseteq R$. 
Applying the same argument for $h_{f_3,u_1} \in H_{f_3u_1}$, 
we see that $ H_{f_3u_1}\cup \{h^{0}_{f_3u_1}\} \subseteq R$. 
Clearly, this shows that $f_3$ and $u_2$ are in $R$ for protection of vertices in 
 $H_{u_1u_2}\cup H_{f_3u_1}$. Applying the same argument for 
$u_2$ and $f_3$, we get  $H_{u_2u_3} \cup \{h^{0}_{u_2u_3}\} \subseteq R$ and 
$H_{f_3u_1}\cup \{h^0_{f_3u_1}\} \subseteq R$, respectively. 
Repeatedly applying the above argument, we get $V_{\triangle}\subseteq R$. 
Observe that this argument can be easily extended to all the vertices of $U$  and $F$. 
Therefore, we see that if $(U\cup F)\cap R \neq \emptyset$, then 
$V_{\triangle}\subseteq R$. \\\\
{\it Case 2:} Suppose $R$ contains $h_{xy}$ from $H_{xy}$ for some $(x,y)\in P$. 
Clearly, this implies $R$ contains both $x$ and $y$ from $U\cup F$. 
Using Case 1, we get $V_{\triangle}\subseteq R$.\\\\
{\it Case 3:} Suppose $R$ contains $ h^0_{xy}$ for some $(x,y)\in P$. 
Clearly, this implies that $H_{xy} \subseteq R$. 
Using Case 2, we get  $V_{\triangle} \subseteq R$.\\\\
Therefore we proved  that if $V_{\triangle} \cap R \neq \emptyset $ then $V_{\triangle} \subseteq R$. 
Next we claim that if $R$ is non-empty then $R$ contains the set 
$V_{\triangle}$. Since $R$ is non-empty,  we see that $R$ must contain a 
vertex from graph $G$. We consider the
following cases:\\\\
{\it Case 1:} Suppose $R$ contains $a^{s}$ from $A_{s}$ for some $s\in S$. Then we know that $d_{R^{c}}(a^{s})\geq |N_{U}(a^{s})|+5$. We see that $a^{s}$ is protected if and only if $ F\cap R \neq \emptyset$. This implies that $V_{\triangle} \cap R \neq \emptyset$ which implies  $V_{\triangle} \subseteq R$.\\\\
{\it Case 2:} Suppose $R$ contains $x_{s}$ for some $s\in S$. We know that $x_{s}$ has at least two neighbours in $A_{s}$ as $\text{max}(s)+1\geq 2$. This implies that $x_{s}$ is protected if and only if at least one vertex $a_{s}\in A_{s}$ is in $R$. 
Now, Case 1 implies that $V_{\triangle}\subseteq R$.\\\\
{\it Case 3:} Suppose $R$ contains $b^{s}\in B_{s}$ for some $s\in S$. 
Then we know that $N(b^{s}) = \{y_{s}\} \cup F$. Clearly, the protection of $b^{s}$ requires at least one vertex from $F$. This implies that $F\cap R \neq \emptyset$. 
Therefore, we have $V_{\triangle}\cap R\neq \emptyset$ and hence $V_{\triangle} \subseteq R$.\\\\
{\it Case 4:} Suppose $R$ contains $y_{s}$ for some $s\in S$. 
We know that $y_{s}$ has at least two neighbours in $B_{s}$ as $\text{max}(s)+1\geq 2$. 
This implies that $y_{s}$ is protected if and only if at least one vertex $b_{s}\in B_{s}$ is in $R$. 
Now, Case 3 implies that $V_{\triangle}\subseteq R$.\\\\
This shows if $R$ is non-empty then $V_{\triangle} \subseteq R$. 
We know  $V_{\triangle}$ contains exactly $ (k+3)N +2k+6 $ many vertices; thus
besides the vertices of $V_{\triangle}$, there are 
at most $\sum\limits_{i=1}^{n}(\text{max}(s_{i})+1)+k'$  vertices in $R$.
Since $f_{1}\in R$ and $d_{V_{\triangle}}(f_{1}) = d_{V_{\square}}(f_{1}) = 2N$, it must have at least $\sum\limits_{i=1}^{n}(\text{max}(s_{i})+1)$ many neighbours 
in $R$ from the set $\bigcup\limits_{s\in S} A_{s}\cup B_{s}$. 
We also observe that if a vertex $a^s$ from the set $A_{s}$ 
is in the solution  then $x_s$ also lie in the solution for the protection of $a^s$. 
This shows that at most $k'$ many sets of the form $A_{s}$  contribute to the solution as 
otherwise the size of solution  exceeds $r$. Therefore, any arbitrary defensive alliance $R$ 
of size at most 
$r$  can be transformed to another defensive alliance $R'$ of size at most $r$ as follows:
$$ R' =  V_{\triangle} \bigcup\limits_{x_{s}\in R} A_{s}\cup \{x_{s}\} \bigcup\limits_{x_{s}\in V(G)\setminus R} B_{s}.$$ 
We define a subset $S'=\Big\{ s \in S ~|~ x_s\in R'\Big\}.$
Clearly, $|S'|\leq k'$. We claim that $\sum\limits_{s\in S'} s(i) \geq t(i)$ for all $1 \leq i \leq k$. 
 Assume for the sake of contradiction that $\sum\limits_{s\in S'} s(i) < t(i)$ for some $i \in \{1,2,\ldots,k\}$. 
Note that  
\begin{equation*}
 d_{R'}(u_i)=\sum\limits_{s \in S'} {s(i)} +2N
\end{equation*}
and 
\begin{equation*}
  d_{R'^c}(u_i)= \sum\limits_{s \in S\backslash S'} s(i) +|V_{u_i}^{\square}|= \sum\limits_{s \in S\backslash S'} s(i) + \sum\limits_{s \in S} s(i) + 2N - 2(\sum\limits_{s\in S} s(i)-t(i))
\end{equation*}
Then, we have 
\begin{equation*}
\begin{split}
 d_{R'^{c}}(u_i)&=2N + \sum\limits_{s \in S\backslash S'} s(i) + \sum\limits_{s \in S} s(i) - 2(\sum\limits_{s\in S} s(i)-t(i))\\
           &=2N -\Big( \sum\limits_{s \in S} s(i) - \sum\limits_{s\in S\backslash S'} s(i)\Big) + 2t(i)  \\
           &=2N -\sum\limits_{s \in S'} s(i) + 2t(i)\\
           &= 2N + \sum\limits_{s\in S'} s(i) + 2\Big(t(i)-\sum\limits_{s\in S'}s(i) \Big)  \\
           &> 2N + \sum\limits_{s\in S'} s(i) = d_{R'}(u_i)
 \end{split}          
\end{equation*}
We also know that $u_{i}\in R'$, which is a contradiction to the fact that $R'$ is a defensive alliance. 
This shows that $I = (k, k', S, t)$ is a yes instance. \qed\\

 Clearly trees of height at most two are trivially acyclic. 
 Moreover, it is easy to verify that such trees have 
 pathwidth \cite{Kloks94} and treedepth \cite{Sparsity} at most two, which implies:
 
\begin{theorem}\rm
 The {\sc Defensive Alliance} problem 
 is W[1]-hard when parameterized by any of the following parameters:
 \begin{itemize}
     \item the feedback vertex set number,
     \item the treewidth and clique width of the input graph,
     \item the pathwidth and treedepth of the input graph,
 \end{itemize}
  even when restricted to bipartite graphs.
\end{theorem}

 \section{No Polynomial Kernel Parameterized by Vertex Cover Number}
 A set $C\subseteq V$ is a vertex cover of $G=(V,E)$ if each edge $e\in E$ has at least one endpoint 
 in $X$. The minimum size of a vertex cover in $G$ is the {\it vertex cover number} of $G$, 
 denoted by $vc(G)$.
 Parameterized by vertex cover number $vc$, the {\sc Defensive Alliance} problem is FPT \cite{KIYOMI201791} 
 and in this section we prove the following 
 kernelization hardness of  the {\sc Defensive Alliance} problem.
 
  \begin{theorem}\label{ppt}\rm
The {\sc  Defensive Alliance}  problem parameterized by the vertex cover number of the input graph does not admit a polynomial 
compression unless coNP $\subseteq$ NP/poly.
 \end{theorem}
 
 \noindent To prove Theorem \ref{ppt}, we give a polynomial parameter transformation (PPT) from the well-known 
 {\sc Red Blue Dominating Set} problem (RBDS) to {\sc Defensive Alliance} parameterized by vertex cover number. 
 Recall that in RBDS we are given a bipartite graph $G=(T\cup S,E)$ and an integer $k$, and we are 
 asked whether there exists a vertex set $X\subseteq S$ of size at most $k$ such that every vertex in $T$
 has at least one neighbour in $X$. We also refer to the vertices of $T$ as {\it terminals} and 
 to the vertices of $S$ as {\it sources} or {\it nonterminals}. The following theorem is known:
 
  \begin{theorem}\label{RBDS} \rm \cite{fomin_lokshtanov_saurabh_zehavi_2019}
 RBDS parameterized by $|T|$ does not admit a polynomial compression unless coNP $\subseteq$ NP/poly.
 \end{theorem}

 \subsection{Proof of Theorem \ref{ppt}} 
\begin{figure}[ht]
    \centering
   \begin{tikzpicture}[scale=0.6]

\node (l1) at (-16, -5) [label=above:$G$]{};
\node (l2) at (-4.5, -7.5) [label=above:$G'$]{};
\node[circle, draw, inner sep=0pt, minimum size=5pt] (t) at (-17,-2) [label=above:$t$] {};
\node[circle, draw, inner sep=0pt, minimum size=5pt] (t') at (-17,-3) [label=below:$t'$] {};
\node[circle, draw, inner sep=0pt, minimum size=5pt] (s) at (-15,-2) [label=above:$s$] {};
\node[circle, draw, inner sep=0pt, minimum size=5pt] (s') at (-15,-3) [label=below:$s'$] {};

  \draw [draw=orange](0,3) ellipse (0.5cm and 1.5cm);
  \draw [draw=orange](0,-3) ellipse (0.5cm and 1.5cm);
  \draw [draw=orange](-10,0) ellipse (0.5cm and 1.5cm);
  
\draw [draw=orange](-6,5) ellipse (1.5cm and 0.5cm);
\draw [draw=orange](-6,-5) ellipse (1.5cm and 0.5cm);

\node[circle,draw, inner sep=0pt, minimum size=5pt] (t1) at (0,4) [label=below:$t_1$] {};
\node[] at (1.5,3.5) [label=below:$T_1$] {};
\node[] at (-1,4.5) [label=above:$V_{t_1}$] {};
\node[circle,draw, inner sep=0pt, minimum size=5pt] (t11) at (-1,4.5) [] {};
\node[circle,draw, inner sep=0pt, minimum size=5pt] (t12) at (-1,4) [] {};
\node[circle,draw, inner sep=0pt, minimum size=5pt] (t13) at (-1,3.3) [] {};
\path 
(t12) edge[dotted] (t13);
\draw [draw=orange](-1,3.9) ellipse (0.4cm and 0.9cm);

\node[circle, draw, inner sep=0pt, minimum size=5pt] (t'1) at (0,2) [label=above:$t'_1$] {};
\node[] at (-1,1.3) [label=below:$V_{t_1^{\prime}}$] {};
\node[circle, draw, inner sep=0pt, minimum size=5pt] (t'11) at (-1,2.5) [] {};
\node[circle, draw, inner sep=0pt, minimum size=5pt] (t'12) at (-1,2) [] {};
\node[circle, draw, inner sep=0pt, minimum size=5pt] (t'13) at (-1,1.3) [] {};
\path 
(t'12) edge[dotted] (t'13);
\draw [draw=orange](-1,1.9) ellipse (0.4cm and 0.9cm);

\node[circle,draw, inner sep=0pt, minimum size=5pt] (t2) at (0,-2) [label=below:$t_2$] {};
\node[] at (1.5,-2.5) [label=below:$T_2$] {};
\node[] at (-1,-1.5) [label=above:$V_{t_2}$] {};
\node[circle,draw, inner sep=0pt, minimum size=5pt] (t21) at (-1,-1.5) [] {};
\node[circle,draw, inner sep=0pt, minimum size=5pt] (t22) at (-1,-2) [] {};
\node[circle,draw, inner sep=0pt, minimum size=5pt] (t23) at (-1,-2.7) [] {};
\path 
(t22) edge [dotted] (t23);
\draw [draw=orange](-1,-2.1) ellipse (0.4cm and 0.9cm);

\node[circle, draw, inner sep=0pt, minimum size=5pt] (t'2) at (0,-4) [label=above:$t'_2$] {};
\node[]() at (-1,-4.7) [label=below:$V_{t_2^{\prime}}$] {};
\node[circle, draw, inner sep=0pt, minimum size=5pt] (t'21) at (-1,-3.5) [] {};
\node[circle, draw, inner sep=0pt, minimum size=5pt] (t'22) at (-1,-4) [] {};
\node[circle, draw, inner sep=0pt, minimum size=5pt] (t'23) at (-1,-4.7) [] {};
\path 
(t'22) edge [dotted] (t'23);
\draw [draw=orange](-1,-4.1) ellipse (0.4cm and 0.9cm);

\node[circle,draw, inner sep=0pt, minimum size=5pt] (t0) at (-10,1) [label=below:$t_0$] {};
\node[] at (-11.5,0.5) [label=below:$T_0$] {};
\node[circle, draw, inner sep=0pt, minimum size=5pt] (t'0) at (-10,-1) [label=above:$t'_0$] {};
\node[circle,draw, inner sep=0pt, minimum size=5pt] (s1) at (-7,5) [label=above:$s_1$] {};
\node[] at (-6,6) [label=above:$S_1$] {};
\node[circle, draw, inner sep=0pt, minimum size=5pt] (s'1) at (-5,5) [label=above:$s'_1$] {};
\node[circle,draw, inner sep=0pt, minimum size=5pt] (s0) at (-7,-5) [label=below:$s_0$] {};
\node[] at (-6,-5.7) [label=below:$S_0$] {};
\node[circle, draw, inner sep=0pt, minimum size=5pt] (s'0) at (-5,-5) [label=below:$s'_0$] {};
\node[circle,draw, inner sep=0pt, minimum size=5pt] (a) at (-6,2) [label=below:$a$] {};
\node[circle,draw, inner sep=0pt, minimum size=5pt] (b) at (-6,0) [label=below:$b$] {};
\node[circle,draw, inner sep=0pt, minimum size=5pt] (c) at (-6,-2) [label=below:$c$] {};

\node[circle, draw, inner sep=0pt, minimum size=5pt] (x*) at (-2,5.5) [label=above:$x^*$] {};
\path
(t) edge [thick, color=red] (s)
(t') edge [thick, color=blue] (s)
(t') edge [thick, color=green] (s')
(t1) edge (t11)
(t1) edge (t12)
(t1) edge (t13)
(t'1) edge (t'11)
(t'1) edge (t'12)
(t'1) edge (t'13)
(t2) edge (t21)
(t2) edge (t22)
(t2) edge (t23)
(t'2) edge (t'21)
(t'2) edge (t'22)
(t'2) edge (t'23)
(a) edge (t11)
(a) edge (t12)
(a) edge (t13)
(a) edge (t'11)
(a) edge (t'12)
(a) edge (t'13)
(b) edge (t11)
(b) edge (t12)
(b) edge (t13)
(b) edge (t'11)
(b) edge (t'12)
(b) edge (t'13)
(c) edge (t11)
(c) edge (t12)
(c) edge (t13)
(c) edge (t'11)
(c) edge (t'12)
(c) edge (t'13)

(a) edge (t21)
(a) edge (t22)
(a) edge (t23)
(a) edge (t'21)
(a) edge (t'22)
(a) edge (t'23)
(b) edge (t21)
(b) edge (t22)
(b) edge (t23)
(b) edge (t'21)
(b) edge (t'22)
(b) edge (t'23)
(c) edge (t21)
(c) edge (t22)
(c) edge (t23)
(c) edge (t'21)
(c) edge (t'22)
(c) edge (t'23)

(a) edge (s1)
(a) edge (s'1)
(a) edge (t0)
(a) edge (t'0)
(b) edge (t0)
(b) edge (t'0)
(t0) edge [thick, color=red] (s0)
(t0) edge [thick, color=red] (s1)
(t1) edge [thick, color=red] (s1)
(t2) edge [thick, color=red] (s0)
(t'0) edge [thick, color=blue] (s0)
(t'0) edge [thick, color=blue] (s1)
(t'1) edge [thick, color=blue] (s1)
(t'2) edge [thick, color=blue] (s0)
(t'0) edge [thick, color=green] (s'0)
(t'0) edge [thick, color=green] (s'1)
(t'1) edge [thick, color=green] (s'1)
(t'2) edge [thick, color=green] (s'0)
(x*) edge (s1)
(x*) edge (s'1)
(x*) edge (t11)
(x*) edge (t12)
;

\end{tikzpicture}

    \caption{ PPT from RBDS to {\sc Defensive Alliance}, where graph $G$ is show on the left and $G'$ is shown on the right. }
    \label{fig:ppt}
\end{figure}
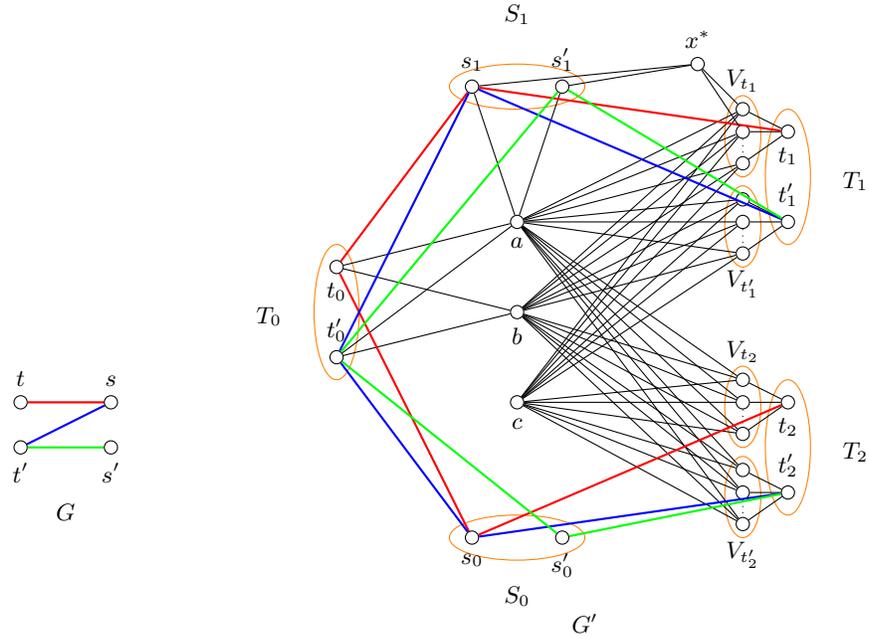

By Theorem \ref{RBDS}, RBDS parameterized by $|T|$ does not admit a polynomial compression unless coNP $\subseteq$ NP/poly.
To prove Theorem \ref{ppt}, we give a PPT from RBDS parameterized by $|T|$ to {\sc Defensive Alliance} parameterized by 
the vertex cover number. Given an instance $(G = (T\cup S, E),k)$ of RBDS, we construct  an instance 
$(G',k')$ of {\sc Defensive Alliance} as follows. 
Take three distinct copies $T_0,T_1,T_2$ of $T$, and let $t_i$ be the copy
of $t\in T$ in $T_i$. Similarly, take two distinct copies $S_0,S_1$ of $S$, and let $s_i$ be the 
copy of $s\in S$  in $S_i$. Now for every vertex in $t \in T_1 \cup T_2$, we introduce a set 
$V_t = \{ t_{1},\ldots,t_{4\ell}\}$ of vertices adjacent to $t$ where the number $\ell$ is 
defined later in the proof. Moreover, create three vertices $a,b$ and $c$. 
The vertices $a,b$ and $c$ are adjacent to all the vertices in 
set $\bigcup\limits_{t\in T_{1} \cup T_{2}}V_t$. We also make  $a$ and $b$ adjacent to every vertex in $T_0$; 
and make $a$ adjacent to every vertex in $S_1$.
If  $(t,s)\in E(G)$ then we add the edges $(t_0,s_0), (t_0,s_1), (t_1,s_1)$ and $(t_2,s_0)$ in $E(G')$.
Finally, we add a vertex $x^*$  which is adjacent to every vertex in $S_1$ and also 
adjacent to exactly $|S|$ many arbitrary vertices from  $V_t$ for some $t\in T_{1} \cup T_{2}$. 
We observe that $C=T_0 \cup T_{1} \cup T_{2} \cup \{a,b,c,x^*\}$ is a vertex cover of $G'$. 
Therefore the vertex cover size of $G'$ is bounded by $3|T|+4$. We set $k' = |T|+|S|+k+1$ and $l=4k'$. 
See Figure \ref{fig:ppt} for an illustration. We now claim that $G$ is a yes-instance of RBDS if and only if 
$G'$ is a yes-instance of  {\sc Defensive Alliance}.

\par Suppose there exists a vertex set $X\subseteq S$ of size at most $k$ in $G$ such that every vertex in
 $T$ has at least one neighbour in $X$. We claim that the set 
 $R = S_1 \cup \big\{ s_0 \in S_0 ~|~s\in X \big \} \cup T_0 \cup \{x^*\}$ is a defensive alliance in graph $G'$. Let $x$ be an arbitrary element of $R$.  We prove that $x$ is protected in $R$.\\
 
\noindent{\it Case 1:} Suppose $x\in S_1$. Note that 
$N_R(x)=N_{T_0}(x)\cup \{x^*\}$. 
Thus, including itself, it has $d_G(x)+2$ defenders in $G'$. 
The attackers of $x$ consist of 
elements of $N_{T_1}(x)$ and element $a$. Hence $x$ has $d_G(x)+1$ attackers. This shows that  $x$ has at least as many defenders
as attackers; hence $x$ is protected. \\

\noindent{\it Case 2:} Suppose  $x\in \big\{ s_0 \in S_0 ~|~s\in X \big \}$. 
Note that 
$N_R(x)= N_{T_0}(x)$. Thus, including itself, it has $d_G(x)+1$ defenders in $G'$. The attackers of $x$ consist of 
elements of $N_{T_2}(x)$. Hence $x$ has $d_G(x)$ attackers in $G'$. This shows that  $x$ is protected.\\

\noindent{\it Case 3:} Suppose $x\in T_0$. Clearly, including itself, $x$ has $2d_G(x)+3$ neighbours in $G'$.  
Thus 
it requires at least $d_G(x)+2$ many defenders in $G'$. Note that, including itself, 
$x$ has $d_G(x)+1$ neighbours in $S_1\subseteq R$. Therefore, it  requires at least one neighbour from the set 
$\big\{ s_0 \in S_0 ~|~s\in X \big \}$ inside the solution and this is true because $G$ is a yes instance. \\

\noindent{\it Case 4:} Suppose $x=x^*$. It has the same number of defenders and attackers in $G'$. This shows
that $x$ is protected.\\

\par Conversely, suppose there exists a defensive alliance $R$ of size at most $k'$ in $G'$. 
We observe that no vertex  from the set
$Q = T_{1} \cup T_{2} \cup \{a,b,c\} \bigcup\limits_{t\in T_{1}\cup T_{2}} V_t $
can be part of $R$ as otherwise its size will exceed $k'$. 
Since $R$ is non-empty, it must contain a vertex from one of the sets $\{x^*\}, 
S_1,T_0$ or $S_0$.\\
\noindent{\it Case 1:} Suppose $x^*\in R$. 
Since $x^*$ has $|S|$ many neighbours in $Q$, 
it implies that all the neighbours of $x^*$ in $S_1$ must be inside 
the solution for protection of $x^*$. 
This implies that $S_1\subseteq R$. 
Let $v$ be an arbitrary vertex in $S_1$. 
Note that $v$ has $d_G(v)$  neighbours in $T_0$, and it has $d_G(v)+1$  neighbours in $Q$. 
For protection of $v$ all the neighbours of $v$ in $T_0$ must be part of the solution. 
This implies that $T_0\subseteq R$ as all the vertices in $S_1$ must be protected. 
Note that till now we have added $|S|+|T|+1$ many vertices 
in the solution. 
Therefore, we can  add at most $k$ vertices to the solution from the set $S_0$ as 
otherwise the solution size will exceed $k'$.  Suppose we add a set $X\subseteq S_0$ 
of size at most $k$ to the solution. Consider the protection of vertices in $T_0$. 
If  $v$ is a vertex  of $T_0$, then it has $d_G(v)$  neighbours in $S_0$ and similarly 
$d_G(v)$  neighbours in $S_1$. Excluding itself, $v$ has $2d_G(v)+2$ neighbours in $G'$.
 Thus it requires at least $d_G(v)+1$ many neighbours inside the solution. 
 We know that $d_G(v)$ neighbours are inside the solution due to the fact that $S_1\subseteq R$. 
 Therefore, it  requires at least one neighbour from $S_0$ inside the solution. 
 Since there exists a set $X\subseteq S$ of size at most $k$ such that all the vertices in $T_0$ are protected,
 it shows that all vertices in $T_0$ have at least one neighbour in $X$. This proves that $G$ is a yes instance.\\
 
\noindent{\it Case 2:} Suppose $R$ contains a vertex $v$  from the set $S_1$. In this case, 
the protection of $v$ requires $x^*$ to be inside the solution and then the same argument as in
Case 1 will lead to the proof. \\

\noindent{\it Case 3:} Suppose $R$ contains a vertex $v$  from the set $T_0$. 
Excluding itself, $v$ has $2d_G(v)+2$ neighbours in $G'$.
Thus it requires at least $d_G(v)+1$ many neighbours from $S_0\cup S_1$ inside the solution.
This implies that at least one neighbour from the set $S_1$ must be inside the solution. 
Now the same argument  as in Case 2 will lead to the proof. \\

\noindent{\it Case 4:} Suppose $R$ contains a vertex $v$  from the set $S_0$. 
Clearly, it has $d_G(v)$ neighbours in $T_{2}\subseteq Q$  and $d_G(v)$ neighbours in  $T_0$. 
Since the vertices in the set $Q$ cannot be part of the solution, the protection of $v$ will 
imply that all the neighbours of $v$ in $T_0$ are part of the solution. 
In other words, there exists a vertex in $T_0$ which is inside the solution. 
Now the same argument as in Case 3 will lead to the proof. \\

\noindent This proves that $G$ is a yes-instance. \qed\\

\section{ {\sc Defensive Alliance} has no Subexponential Algorithm}
In this section, we prove lower bound based on ETH
for the time needed to solve the {\sc Defensive Alliance} problem. 
In order to prove that a too fast algorithm for {\sc Defensive Alliance} contradicts ETH, we give a reduction from {\sc Vertex Cover} in graphs of maximum degree 3
 and argue that a too fast algorithm for 
{\sc Defensive Alliance} would solve {\sc Vertex Cover} in graphs of maximum degree 3
 in time $2^{o(n)}$.  Johnson  and Szegedy \cite{soda1999} proved that, assuming ETH, there is no algorithm with running time $2^{o(n)}$ to compute a minimum vertex cover in graphs of maximum degree 3.

\begin{figure}
    \centering
    \includegraphics[scale=0.15]{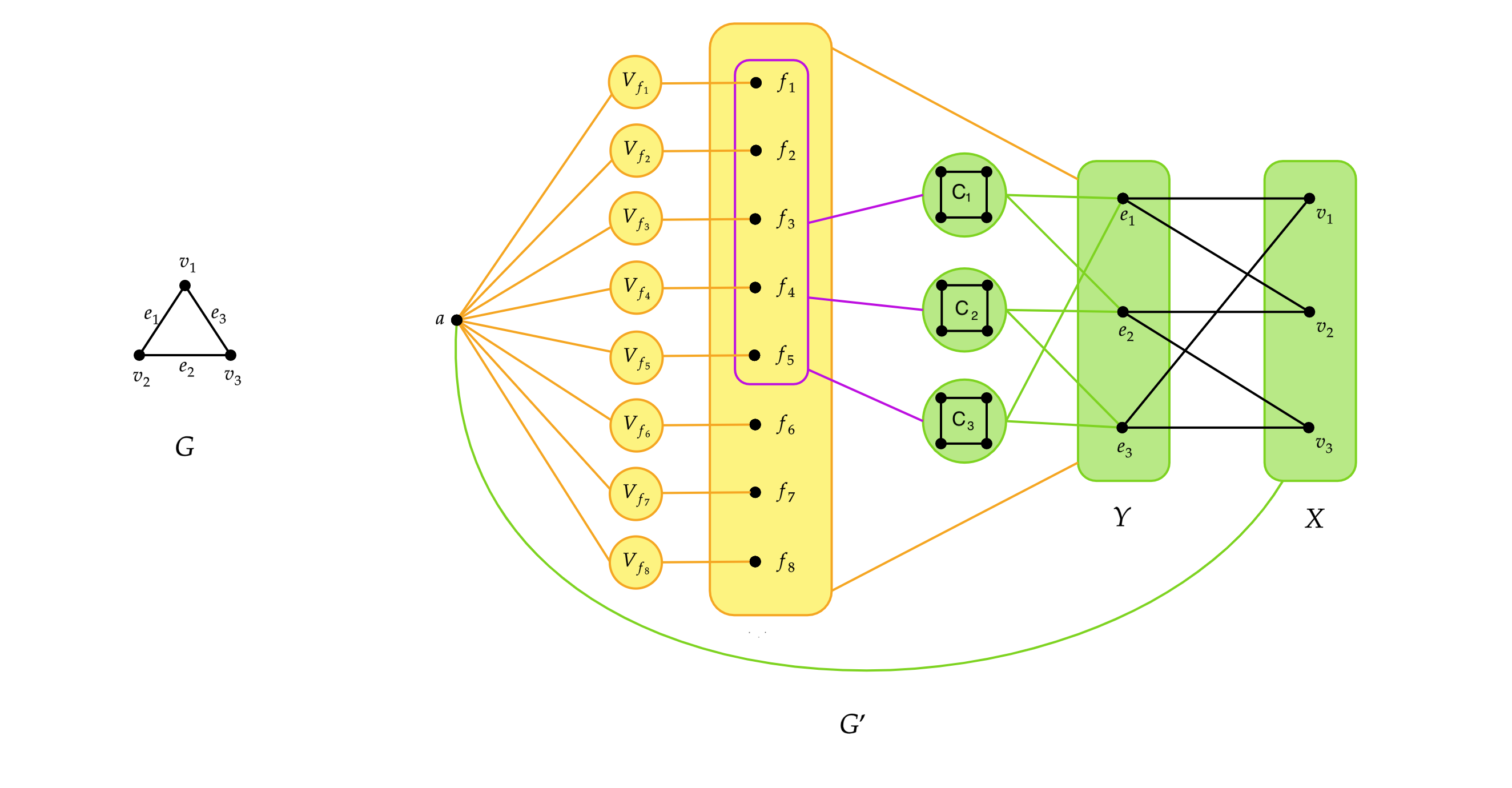}
    \caption{The reduction from {\sc Vertex Cover} to {\sc Defensive Alliance}. }
    \label{fig:ETH}
\end{figure}

\begin{theorem}\label{ETH} \rm
 Unless ETH fails, {\sc  Defensive Alliance}  does not admit a $2^{o(n)}$ algorithm where $n$ is the number of vertices  of the input graph.
 \end{theorem}
\proof We give a linear reduction from {\sc Vertex Cover} in graphs of maximum degree 3  to {\sc Defensive Alliance}, that is, a 
polynomial-time algorithm that takes an instance $(G,k)$ of {\sc Vertex Cover}, where
$G$ has $n$ vertices and $m=O(n)$ edges, and 
outputs an equivalent instance of {\sc Defensive Alliance} whose size is bounded by $O(n)$.
 We construct an equivalent instance $(G',k')$
of {\sc Defensive Alliance} the following way. See Figure \ref{fig:ETH}. 
\begin{enumerate}
    \item We introduce the  vertex sets $X$ and $Y$
into $G'$, where
$X=V(G)=\{v_1,\ldots,v_n\}$ and $Y=E(G)=\{e_1,e_2,\ldots,e_m\}$, the edge set of $G$. 
We make $v_i$ adjacent to $e_j$ if and only if $v_i$ is an endpoint of $e_j$. 
\item For every $1\leq i\leq m$, we introduce a cycle $C_i$ of length $4$.
For every $1\leq i\leq m-1$, make every vertex of 
$C_i$ adjacent to $e_i$ and $e_{i+1}$; and make every vertex of $C_m$ adjacent
to $e_m$ and $e_1$.
\item  We add a set $F=\{f_{1},f_{2},\ldots,f_{8}\}$ 
of $8$ new vertices into $G'$. Set $k'=5m+k$. For every vertex $f \in F$ 
 we introduce a set $V_f$ of  $4k'$ new vertices into $G'$ and make them adjacent to $f$. We  make every vertex of
 $\{f_1,f_2,f_3,f_4,f_5\}$ adjacent to every vertex of $C_i$ for $i=1,2,\ldots,m$.
 We also make every vertex of $F$ adjacent to every vertex of $Y$. 
 \item  Finally, we introduce a vertex $a$ and make it adjacent to every vertex of $X\cup \bigcup\limits_{f\in F}{V_f}$.
\end{enumerate}

 We now argue equivalence of the instances. Suppose there exists a vertex cover $S$ of size at most $k$ in $G$. We show that  $D=S \cup Y \bigcup\limits_{i=1}^{m} V(C_i) $ is a 
defensive alliance of size at most $k'$ in $G'$. It is easy to verify that all the vertices in $D$ 
are protected. 

 \par To prove the reverse direction of the equivalence, suppose now that 
 $D$ is a defensive alliance of size at most $k'$ in $G'$. Observe that no vertex from 
$Q = \{a\}\cup F\cup \bigcup\limits_{f\in F} V_f $
can be part of $D$ as otherwise the size of $D$ will exceed $k'$.
To prove this theorem we need the following simple claim:\\
\begin{claim}
Every defensive alliance $D$ of $G'$ contains the set  $Y \bigcup\limits_{i=1}^{m} V(C_i)$.
\end{claim}
\proof 
Since the defensive alliance is non-empty, it must contain a vertex from $X\cup Y \bigcup\limits_{i=1}^{m} V(C_i)$.
 \\\\
\noindent {\it Case 1:} Suppose  $D$ contains  $e_i$ from $Y$. 
We observe that $\text{deg}_{G'}(e_i)=18$.  Note that  eight neighbours of $e_i$  in 
$F$ cannot be part of the solution as they belong to the forbidden set $Q$.
This implies that we need to add at least one vertex 
from the set $V(C_{i-1})\cup V(C_{i})$ to the solution for the protection
of $e_i$. Without loss of generality, suppose we include one vertex 
from $V(C_i)$ in the solution. 
Inclusion of one vertex from the set $V(C_{i})$ in the solution forces 
$V(C_{i})\subseteq D$. This in turn forces $e_{i+1}$ in the solution. Repeatedly applying the above argument, 
we see that $Y \bigcup\limits_{i=1}^{m} V(C_i) \subseteq D$. \\\\
\noindent{\it Case 2:} Suppose $D$ contains an  arbitrary vertex from the set 
$X \bigcup\limits_{i=1}^{m} V(C_i)$.   Then the protection 
of that vertex forces at least one vertex from $Y$ in the solution. 
Using the argument in Case 1, it implies that $Y \bigcup\limits_{i=1}^{m} V(C_i) \subseteq D$.~~~~~~~~~~~~~~~~~~$\lrcorner$

We observe that for every vertex $e\in Y$, we have included eight out of its 
18 neighbours in the solution. 
For the protection of $e$, we  need to include at least one more of its neighbours from 
$X$ in the solution. As we have already added $5m$ vertices 
in the solution, we can add a set $S \subseteq X$ of at most $k$ vertices 
in $D$ such that every 
vertex $e\in Y$ has at least one neighbour in $S$. If such a set $S$ exists then  it forms 
a vertex cover of $G$. This shows that $I$ is a yes instance. \qed  \\

\section{Defensive Alliance on Circle Graphs}
A {\it circle graph} is the intersection graph of a set of chords of a circle. That is, it is an undirected graph whose vertices can be associated with chords of a circle such that two vertices are adjacent if and only if the corresponding chords cross each other.
Here, we prove that the {\sc Defensive Alliance} problem is NP-complete even when restricted to circle graphs, via a reduction from {\sc Dominating Set}.  It is known  that the {\sc Dominating Set} problem  on circle graphs is NP-hard \cite{KEIL199351}.

\begin{theorem}\label{circleTheorem}\rm
The {\sc Defensive Alliance} problem on circle graphs is NP-complete.
\end{theorem}
\noindent On the way towards this result, we provide hardness result for a variant of the  {\sc Defensive Alliance} problem which  we require in the proof of Theorem  \ref{circleTheorem}. The problem {\sc Defensive Alliance$^{\mbox{F}}$} generalizes {\sc Defensive Alliance} where some vertices are forced to be outside the solution; these vertices are called forbidden vertices.  This variant can be formalized as
follows:\\ 

\noindent 
\vspace{3mm}
    \fbox
    {\begin{minipage}{33.7em}\label{SP2}
       {\sc Defensive Alliance$^{\mbox{F}}$}\\
        \noindent{\bf Input:} An undirected graph $G=(V,E)$, a positive integer $r$ and
        a set $V_{\square}\subseteq V(G)$ of forbidden vertices. \\
        \noindent{\bf Question:} Is there a defensive alliance $S\subseteq V$ such that  $1\leq |S|\leq r$, and 
     $S\cap V_{\square}=\emptyset$?
    \end{minipage} }\\
\begin{lemma}\label{FNvds1}\rm
The {\sc Defensive Alliance$^{\mbox{F}}$} problem on circle graphs is NP-complete.
\end{lemma}
\proof It is easy to see  that the problem is in NP. 
To show that the problem is NP-hard we give a polynomial reduction from 
{\sc Dominating Set} on circle graphs. Let $(G,k)$ be an instance of {\sc Dominating Set},   where $G$ is a circle graph. Suppose we are also given the circle  representation $C$ of $G$. Without loss of generality, we can assume that none of the endpoints of chords overlap with each other. We create a 
graph $G'$  and output the instance $(G',V_{\square},k')$. See Figure \ref{circleNP}. The steps given below describe the construction of $G'$:
\begin{itemize}
\item{\bf Step 1:} Take two distinct copies $G_1$ and $G_2$ of $G$ and let $v_i$ be the copy of $v\in V(G)$ in graph $G_i$. For each $v\in V$, make $v_1$ adjacent to every vertex of
$N_{G_2}(v_{2}) \cup \{v_2\}$ and similarly make $v_2$ adjacent to every vertex of
$N_{G_1}(v_1) \cup \{v_1\}$. Note that this operation can be easily incorporated in 
the circle representation by replacing the chord corresponds to $v$ with two
crossing cords correspond to $v_1$ and $v_2$ as shown in Figure \ref{operation1}. 

\begin{figure}[ht]
\begin{center}
\begin{tikzpicture}[scale=0.35]
\node[fill=black, circle, draw=black, inner sep=0, minimum size=0.1cm](a) at (-7,1.5) [label=above:$\color{red} { a}$] {};
\node[fill=black, circle, draw=black, inner sep=0, minimum size=0.1cm](b) at (-8.5,-1.5) [label=left:$\color{blue} {b}$] {};
\node[fill=black, circle, draw=black, inner sep=0, minimum size=0.1cm](c) at (-5.5,-1.5) [label=right:$\color{green} {c}$] {};

\node[fill=black, circle, draw=black, inner sep=0, minimum size=0.1cm](a1) at (-8,-4.5) [label=left:$\color{red} { a_{1}}$] {};
\node[fill=black, circle, draw=black, inner sep=0, minimum size=0.1cm](b1) at (-8,-6) [label=left:$\color{blue} {b_{1}}$] {};
\node[fill=black, circle, draw=black, inner sep=0, minimum size=0.1cm](c1) at (-8,-7.5) [label=left:$\color{green} {c_{1}}$] {};

\node[fill=black, circle, draw=black, inner sep=0, minimum size=0.1cm](a2) at (-6,-4.5) [label=right:$\color{red} { a_{2}}$] {};
\node[fill=black, circle, draw=black, inner sep=0, minimum size=0.1cm](b2) at (-6,-6) [label=right:$\color{blue} {b_{2}}$] {};
\node[fill=black, circle, draw=black, inner sep=0, minimum size=0.1cm](c2) at (-6,-7.5) [label=right:$\color{green} {c_{2}}$] {};

\draw(a)--(b);
\draw(b)--(c);
\draw(a)--(c);
\draw(a1)--(b1);
\draw(b1)--(c1);
\draw(a2)--(b2);
\draw(b2)--(c2);
\draw(a2)--(c2);
\draw(a1)--(b2);
\draw(b1)--(c2);
\draw(a1)--(c2);
\draw(a2)--(b1);
\draw(b2)--(c1);
\draw(a2)--(c1);
\draw(a1)--(a2);
\draw(b1)--(b2);
\draw(c1)--(c2);

\draw (a1).. controls (-8.4,-6) .. (c1);
\draw (a2).. controls (-5.6,-6) .. (c2);

\draw[thick] (0,-6) circle [radius=2]; 

\node[fill=black, circle, draw=black, inner sep=0, minimum size=0.1cm](q01) at (1.732,-5) [] {};
\node[fill=black, circle, draw=black, inner sep=0, minimum size=0.1cm](q02) at (-1.732,-5) [] {};
\node[fill=black, circle, draw=black, inner sep=0, minimum size=0.1cm](q05) at (1.322, -4.5) [] {};
\node[fill=black, circle, draw=black, inner sep=0, minimum size=0.1cm](q06) at (1.322,-7.5) [] {};
\node[fill=black, circle, draw=black, inner sep=0, minimum size=0.1cm](q07) at (-1.322, -4.5) [] {};
\node[fill=black, circle, draw=black, inner sep=0, minimum size=0.1cm](q08) at (-1.322, -7.5) [] {};

\node[fill=black, circle, draw=black, inner sep=0, minimum size=0.1cm](q11) at (1.932,-5.4) [] {};
\node[fill=black, circle, draw=black, inner sep=0, minimum size=0.1cm](q12) at (-1.6,-4.7) [] {};
\node[fill=black, circle, draw=black, inner sep=0, minimum size=0.1cm](q15) at (1, -4.2) [] {};
\node[fill=black, circle, draw=black, inner sep=0, minimum size=0.1cm](q16) at (1,-7.75) [] {};
\node[fill=black, circle, draw=black, inner sep=0, minimum size=0.1cm](q17) at (-1, -4.3) [] {};
\node[fill=black, circle, draw=black, inner sep=0, minimum size=0.1cm](q18) at (-1.522, -7.3) [] {};

\draw[thick, blue](q01)--(q02);
\draw[thick, red](q15)--(q08);
\draw[thick, green](q06)--(q07);
\draw[thick, blue](q11)--(q12);
\draw[thick, red](q05)--(q18);
\draw[thick, green](q16)--(q17);

\draw[thick] (0,0) circle [radius=2];

\node[fill=black, circle, draw=black, inner sep=0, minimum size=0.1cm](q1) at (1.732,1) [] {};
\node[fill=black, circle, draw=black, inner sep=0, minimum size=0.1cm](q2) at (-1.732,1) [] {};
\node(k1) at (0,1) [label=above:$\color{blue} { b}$] {};
\node(k1) at (0.8,0.2) [label=above:$\color{red} {a}$] {};
\node(k1) at (0.8,-0.8) [label=above:$\color{green} {c}$] {};

\node[fill=red, circle, draw=black, inner sep=0, minimum size=0.1cm](q5) at (1.322, 1.5) [] {};
\node[fill=black, circle, draw=black, inner sep=0, minimum size=0.1cm](q6) at (1.322,-1.5) [] {};
\node[fill=black, circle, draw=black, inner sep=0, minimum size=0.1cm](q7) at (-1.322, 1.5) [] {};
\node[fill=black, circle, draw=black, inner sep=0, minimum size=0.1cm](q8) at (-1.322, -1.5) [] {};

\draw[thick, blue](q1)--(q2);
\draw[thick, red](q5)--(q8);
\draw[thick, green](q6)--(q7);

\node (x0) at (-3.5, -3.5) [label=above:$(i)$] {};
\node (x0) at (-3.5, -9.5) [label=above:$(ii)$] {};
\end{tikzpicture}
\caption{ (i) Graph $G$ and its circle representation. (ii) The graph produced after 
the first step of reduction and its circle representation.}
\label{operation1}
\end{center}
 \end{figure}
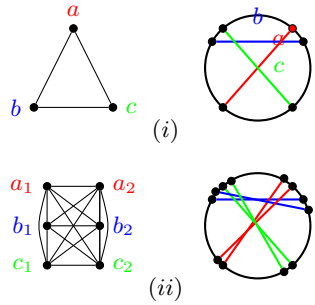
\item{\bf Step 2:} For every $v\in V$, create two sets of vertices $X^v=\{x_1^v,\ldots,x^v_{2n+1}\}$
and $Y^v=\{y_1^v,\ldots,y^v_{2n+1}\}$ and make $v_1,v_2$ adjacent to
every vertex of $X^v\cup Y^v$. This can be easily incorporated in circle representation  by introducing
$2n+1$ parallel chords for the vertices $x_1^v,\ldots, x_{2n+1}^v$  which cross the  
chords for $v_1,v_2$. Similarly, introduce $2n+1$ parallel chords for the vertices $y_1^v,\ldots, y_{2n+1}^v$  which cross the  
chords for $v_1,v_2$, as shown in Figure \ref{operation2}.
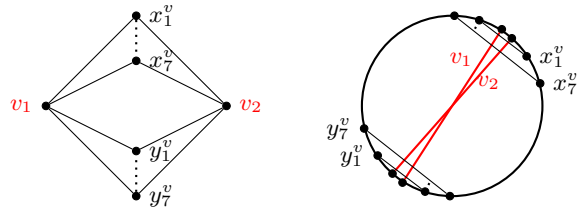
\begin{figure}
\centering
\begin{tikzpicture}[scale=0.6]
\draw[thick] (0,0) circle [radius=2]; 

\node(k1) at (0.2,0.5) [label=above:$\color{red} {v_1}$] {};
\node(k2) at (0.8,0) [label=above:$\color{red} {v_2}$] {};

\node[fill=black, circle, draw=black, inner sep=0, minimum size=0.1cm](q1) at (1.322, 1.5) [] {};
\node[fill=black, circle, draw=black, inner sep=0, minimum size=0.1cm](q2) at (-1.322, -1.5) [] {};

\node[fill=black, circle, draw=black, inner sep=0, minimum size=0.1cm](q3) at (1.1, 1.7) [] {};
\node[fill=black, circle, draw=black, inner sep=0, minimum size=0.1cm](q4) at (-1.1, -1.7) [] {};

\node[fill=black, circle, draw=black, inner sep=0, minimum size=0.1cm](c1) at (1.65, 1.1) [label=right:$x^{v}_{1}$] {};
\node[fill=black, circle, draw=black, inner sep=0, minimum size=0.1cm](c2) at (1.95, 0.5) [label=right:$x^{v}_{7}$] {};

\node[fill=black, circle, draw=black, inner sep=0, minimum size=0.1cm](c3) at (0.6, 1.9) [] {};
\node[fill=black, circle, draw=black, inner sep=0, minimum size=0.1cm](c4) at (0.05,2) [] {};

\node[fill=black, circle, draw=black, inner sep=0, minimum size=0.1cm](c5) at (-1.65, -1.1) [label=left:$y^{v}_{1}$] {};
\node[fill=black, circle, draw=black, inner sep=0, minimum size=0.1cm](c6) at (-1.95, -0.5) [label=left:$y^{v}_{7}$] {};

\node[fill=black, circle, draw=black, inner sep=0, minimum size=0.1cm](c7) at (-0.6,-1.9) [] {};
\node[fill=black, circle, draw=black, inner sep=0, minimum size=0.1cm](c8) at (-0.05,-2) [] {};

\draw[thick, red](q1)--(q2);
\draw[thick, red](q3)--(q4);

\draw(c1)--(c3);
\draw(c2)--(c4);
\draw(c5)--(c7);
\draw(c6)--(c8);

\draw[dotted, thick] (0.6,1.9)--(0.5,1.7);
\draw[dotted, thick] (-0.6,-1.9)--(-0.5,-1.7);

\node[fill=black, circle, draw=black, inner sep=0, minimum size=0.1cm](a1) at (-9, 0) [label=left:$\color{red} v_{1}$] {};
\node[fill=black, circle, draw=black, inner sep=0, minimum size=0.1cm](a2) at (-5, 0) [label=right:$\color{red} v_{2}$] {};

\node[fill=black, circle, draw=black, inner sep=0, minimum size=0.1cm](x1) at (-7,2) [label=right:$x_{1}^{v}$] {};
\node[fill=black, circle, draw=black, inner sep=0, minimum size=0.1cm](x7) at (-7,1) [label=right:$x_{7}^{v}$] {};

\node[fill=black, circle, draw=black, inner sep=0, minimum size=0.1cm](y1) at (-7,-2) [label=right:$y_{7}^{v}$] {};
\node[fill=black, circle, draw=black, inner sep=0, minimum size=0.1cm](y7) at (-7,-1) [label=right:$y_{1}^{v}$] {};

\draw(a1)--(x1);
\draw(a1)--(x7);
\draw(a1)--(y1);
\draw(a1)--(y7);

\draw(a2)--(x1);
\draw(a2)--(x7);
\draw(a2)--(y1);
\draw(a2)--(y7);

\draw[dotted, thick](x1)--(x7);
\draw[dotted, thick](y1)--(y7);

\end{tikzpicture}
     \caption{Illustration of Step 2. Here $2n+1=7$.}
     \label{operation2}
 \end{figure}

\item{\bf Step 3:} For each $x^v\in X^v$, create two 3-vertex cliques $C^1_{x^v}$ and $C^2_{x^v}$, and make  $x^v$ adjacent to every vertex of 
$C^1_{x^v}$ and $C^2_{x^v}$. For $1\leq i\leq 2n$,  make every vertex of  
$C^1_{x_i^v}$ adjacent to every
vertex of $C^1_{x_{i+1}^v}$.
Similarly, make every vertex of $C^2_{x_i^v}$ adjacent to every vertex of $C^2_{x_{i+1}^v}$ for $1\leq i\leq 2n$.
For each $y^v\in Y^v$, create two 3-vertex cliques $C^1_{y^v}$ and $C^2_{y^v}$, and make $y^v$ adjacent to every vertex of $C^1_{y^v}$ and $C^2_{y^v}$. 
Make every vertex of $C^1_{y_i^v}$ adjacent to every vertex of $C^1_{y_{i+1}^v}$ for $1\leq i\leq 2n$.
Similarly, make every vertex of $C^2_{y_i^v}$ adjacent to every vertex of $C^2_{y_{i+1}^v}$ for $1\leq i\leq 2n$.
We start at an arbitrary
 vertex on  the circle representation of $C$ of $G$ and then traverse the circle in counter clockwise direction. 
 We record the sequence in which the chords are visited. 
 For example, in Figure \ref{operation1}(i), if we start at
 the red vertex on the circle, 
 then the sequence in which the chords are visited, is $a,b,c, a,b,c$.
 Note that every vertex  appears twice in the sequence as every chord 
 is visited twice while traversing the circle. Thus we get a sequence $S$ of 
 length $2n$ where
 $n$ is the number of chords. We use the sequence to connect  newly added
 cliques.
 For every consecutive pair $(u,v)$ in the sequence $S$, make every vertex 
 of $ C^2_{x_{2n+1}^u} $ adjacent to every vertex
 of $C^1_{x_{2n+1}^v}$ when both $u$ and $v$ appear for the first time in the sequence
 $S$;
 make every vertex of $C^2_{y_{2n+1}^u} $ adjacent to every vertex 
 of $C^1_{y_{2n+1}^v}$ when both $u$ and $v$ appear for the second time;
 and make every vertex of $ C^2_{x_{2n+1}^u} $ adjacent to every vertex 
 of $C^1_{y_{2n+1}^v}$ when $u$ appears for the first time and $v$ appear for the second time. These adjacency  are shown in green color in  Figure \ref{circleNP} and \ref{step3}.
 
 \begin{figure}
\centering
\begin{tikzpicture}[scale=0.6]

\node[fill=black, circle, draw=black, inner sep=0, minimum size=0.1cm](c1) at (0.4, 0) [label=left:$\color{green} c_{1}$] {};
\node[fill=black, circle, draw=black, inner sep=0, minimum size=0.1cm](c2) at (5.6, 0) [label=right:$\color{green} c_{2}$] {};

\node[fill=black, circle, draw=black, inner sep=0, minimum size=0.1cm](x1) at (3,3.6) [label=above:$x_{1}^{c}$] {};
\node[fill=black, circle, draw=black, inner sep=0, minimum size=0.1cm](1x11) at (6,3.5) [] {};
\node[fill=black, circle, draw=black, inner sep=0, minimum size=0.1cm](1x12) at (6.4,3.5) [] {};
\node[fill=black, circle, draw=black, inner sep=0, minimum size=0.1cm](1x13) at (6.2,3.9) [] {};
\node[ circle, draw=orange, thick, inner sep=0, minimum size=0.52cm](x1c2) at (6.2,3.7) [label=right:$C^2_{x_1^c}$] {};

\path
(1x11) edge (1x12)
(1x12) edge (1x13)
(1x13) edge (1x11)
(x1) edge [color=orange,thick] (x1c2);
\node[fill=black, circle, draw=black, inner sep=0, minimum size=0.1cm](2x11) at (-0.4,3.5) [] {};
\node[fill=black, circle, draw=black, inner sep=0, minimum size=0.1cm](2x12) at (0,3.5) [] {};
\node[fill=black, circle, draw=black, inner sep=0, minimum size=0.1cm](2x13) at (-0.2,3.9) [] {};
\node[ circle, draw=orange, thick, inner sep=0, minimum size=0.52cm](x1c1) at (-0.2,3.7) [label=left:$C^1_{x_1^c}$] {};

\path
(2x11) edge (2x12)
(2x12) edge (2x13)
(2x13) edge (2x11)
(x1) edge [color=orange,thick] (x1c1);

\node[fill=black, circle, draw=black, inner sep=0, minimum size=0.1cm](x2) at (3,2.6) [label=above:$x_{2}^{c}$] {};
\node[fill=black, circle, draw=black, inner sep=0, minimum size=0.1cm](1x21) at (6,2.5) [] {};
\node[fill=black, circle, draw=black, inner sep=0, minimum size=0.1cm](1x22) at (6.4,2.5) [] {};
\node[fill=black, circle, draw=black, inner sep=0, minimum size=0.1cm](1x23) at (6.2,2.9) [] {};
\node[ circle, draw=orange, thick, inner sep=0, minimum size=0.52cm](x2c2) at (6.2,2.7) [label=right:$C^2_{x_2^c}$] {};
\path
(1x21) edge (1x22)
(1x22) edge (1x23)
(1x23) edge (1x21)
(x2) edge [color=orange,thick] (x2c2);
\node[fill=black, circle, draw=black, inner sep=0, minimum size=0.1cm](2x21) at (-0.4,2.5) [] {};
\node[fill=black, circle, draw=black, inner sep=0, minimum size=0.1cm](2x22) at (0,2.5) [] {};
\node[fill=black, circle, draw=black, inner sep=0, minimum size=0.1cm](2x23) at (-0.2,2.9) [] {};
\node[ circle, draw=orange, thick, inner sep=0, minimum size=0.52cm](x2c1) at (-0.2,2.7) 
[label=left:$C^1_{x_2^c}$] {};
\path
(2x21) edge (2x22)
(2x22) edge (2x23)
(2x23) edge (2x21)
(x2) edge [color=orange,thick] (x2c1);

\node[fill=black, circle, draw=black, inner sep=0, minimum size=0.1cm](x7) at (3,1.2) [label=right:$x_{7}^{c}$] {};
\node[fill=black, circle, draw=black, inner sep=0, minimum size=0.1cm](1x71) at (6,1) [] {};
\node[fill=black, circle, draw=black, inner sep=0, minimum size=0.1cm](1x72) at (6.4,1) [] {};
\node[fill=black, circle, draw=black, inner sep=0, minimum size=0.1cm](1x73) at (6.2,1.4) [] {};
\node[ circle, draw=orange, thick, inner sep=0, minimum size=0.52cm](x7c2) at (6.2,1.2) 
[label=right:$C^2_{x_7^c}$] {};
\path
(1x71) edge (1x72)
(1x72) edge (1x73)
(1x73) edge (1x71)
(x7) edge [color=orange,thick] (x7c2);
\node[fill=black, circle, draw=black, inner sep=0, minimum size=0.1cm](2x71) at (-0.4,1) [] {};
\node[fill=black, circle, draw=black, inner sep=0, minimum size=0.1cm](2x72) at (0,1) [] {};
\node[fill=black, circle, draw=black, inner sep=0, minimum size=0.1cm](2x73) at (-0.2,1.4) [] {};
\node[ circle, draw=orange, thick, inner sep=0, minimum size=0.52cm](x7c1) at (-0.2,1.2) 
[label=left:$C^1_{x_7^c}$]{};
\path
(2x71) edge (2x72)
(2x72) edge (2x73)
(2x73) edge (2x71)
(x7) edge [color=orange,thick] (x7c1);

\node[fill=black, circle, draw=black, inner sep=0, minimum size=0.1cm](y1) at (3,-1) [label=right:$y_{1}^{c}$] {};
\node[fill=black, circle, draw=black, inner sep=0, minimum size=0.1cm](1y11) at (6,-1) [] {};
\node[fill=black, circle, draw=black, inner sep=0, minimum size=0.1cm](1y12) at (6.4,-1) [] {};
\node[fill=black, circle, draw=black, inner sep=0, minimum size=0.1cm](1y13) at (6.2,-0.6) [] {};
\node[ circle, draw=orange, thick, inner sep=0, minimum size=0.52cm](y1c2) at (6.2,-0.8) [label=right:$C^2_{y_1^c}$] {};
\path
(1y11) edge (1y12)
(1y12) edge (1y13)
(1y13) edge (1y11)
(y1) edge [color=orange,thick] (y1c2);

\node[fill=black, circle, draw=black, inner sep=0, minimum size=0.1cm](2y11) at (-0.4,-1) [] {};
\node[fill=black, circle, draw=black, inner sep=0, minimum size=0.1cm](2y12) at (0,-1) [] {};
\node[fill=black, circle, draw=black, inner sep=0, minimum size=0.1cm](2y13) at (-0.2,-0.6) [] {};
\node[ circle, draw=orange, thick, inner sep=0, minimum size=0.52cm](y1c1) at (-0.2,-0.8) [label=left:$C^1_{y_1^c}$] {};
\path
(2y11) edge (2y12)
(2y12) edge (2y13)
(2y13) edge (2y11)
(y1) edge [color=orange,thick] (y1c1);
(y1) edge [color=orange,thick] (y1c2);

\node[fill=black, circle, draw=black, inner sep=0, minimum size=0.1cm](y2) at (3,-2) [label=right:$y_{2}^{c}$] {};
\node[fill=black, circle, draw=black, inner sep=0, minimum size=0.1cm](1y21) at (6,-2) [] {};
\node[fill=black, circle, draw=black, inner sep=0, minimum size=0.1cm](1y22) at (6.4,-2) [] {};
\node[fill=black, circle, draw=black, inner sep=0, minimum size=0.1cm](1y23) at (6.2,-1.6) [] {};
\node[ circle, draw=orange, thick, inner sep=0, minimum size=0.52cm](y2c2) at (6.2,-1.8) [label=right:$C^2_{y_2^c}$] {};
\path
(1y21) edge (1y22)
(1y22) edge (1y23)
(1y23) edge (1y21);
\node[fill=black, circle, draw=black, inner sep=0, minimum size=0.1cm](2y21) at (-0.4,-2) [] {};
\node[fill=black, circle, draw=black, inner sep=0, minimum size=0.1cm](2y22) at (0,-2) [] {};
\node[fill=black, circle, draw=black, inner sep=0, minimum size=0.1cm](2y23) at (-0.2,-1.6) [] {};
\node[ circle, draw=orange, thick, inner sep=0, minimum size=0.52cm](y2c1) at (-0.2,-1.8) [label=left:$C^1_{y_2^c}$] {};
\path
(2y21) edge (2y22)
(2y22) edge (2y23)
(2y23) edge (2y21)
(y2) edge [color=orange,thick] (y2c1)
(y2) edge [color=orange,thick] (y2c2);

\node[fill=black, circle, draw=black, inner sep=0, minimum size=0.1cm](y7) at (3,-3.2) [label=right:$y_{7}^{c}$] {};
\node[fill=black, inner sep=0, minimum size=0] at (3,-4) [label=below:$G'$] {};

\node[fill=black, circle, draw=black, inner sep=0, minimum size=0.1cm](1y71) at (6,-3.2) [] {};
\node[fill=black, circle, draw=black, inner sep=0, minimum size=0.1cm](1y72) at (6.4,-3.2) [] {};
\node[fill=black, circle, draw=black, inner sep=0, minimum size=0.1cm](1y73) at (6.2,-2.8) [] {};
\node[ circle, draw=orange, thick, inner sep=0, minimum size=0.52cm](y7c2) at (6.2,-3) [label=right:$C^2_{y_7^c}$] {};
\path
(1y71) edge (1y72)
(1y72) edge (1y73)
(1y73) edge (1y71);
\node[fill=black, circle, draw=black, inner sep=0, minimum size=0.1cm](2y71) at (-0.4,-3.2) [] {};
\node[fill=black, circle, draw=black, inner sep=0, minimum size=0.1cm](2y72) at (0,-3.2) [] {};
\node[fill=black, circle, draw=black, inner sep=0, minimum size=0.1cm](2y73) at (-0.2,-2.8) [] {};
\node[ circle, draw=orange, thick, inner sep=0, minimum size=0.52cm](y7c1) at (-0.2,-3) [label=left:$C^1_{y_7^c}$] {};
\path
(2y71) edge (2y72)
(2y72) edge (2y73)
(2y73) edge (2y71)
(y7) edge [color=orange,thick] (y7c1)
(y7) edge [color=orange,thick] (y7c2);

\draw(c1)--(x1);
\draw(c1)--(x2);
\draw(c1)--(x7);
\draw(c1)--(y1);
\draw(c1)--(y2);
\draw(c1)--(y7);

\draw(c2)--(x1);
\draw(c2)--(x2);
\draw(c2)--(x7);
\draw(c2)--(y1);
\draw(c2)--(y2);
\draw(c2)--(y7);

\draw[dotted](x2)--(x7);
\draw[dotted](y2)--(y7);

\node[fill=black, circle, draw=black, inner sep=0, minimum size=0.1cm](b1) at (1.2, 9) [label=left:$\color{blue} b_{1}$] {};
\node[fill=black, circle, draw=black, inner sep=0, minimum size=0.1cm](b2) at (4.8, 9) [label=right:$\color{blue} b_{2}$] {};

\node[fill=black, circle, draw=black, inner sep=0, minimum size=0.1cm](bx1) at (3,12.2) [label=right:$x_{1}^{b}$] {};
\node[fill=black, circle, draw=black, inner sep=0, minimum size=0.1cm](b1x11) at (6,12.5) [] {};
\node[fill=black, circle, draw=black, inner sep=0, minimum size=0.1cm](b1x12) at (6.4,12.5) [] {};
\node[fill=black, circle, draw=black, inner sep=0, minimum size=0.1cm](b1x13) at (6.2,12.9) [] {};
\node[ circle, draw=orange, thick, inner sep=0, minimum size=0.52cm](b1c2) at (6.2,12.7) 
[label=right:$C^2_{x_1^b}$] {};
\path
(b1x11) edge (b1x12)
(b1x12) edge (b1x13)
(b1x13) edge (b1x11);
\node[fill=black, circle, draw=black, inner sep=0, minimum size=0.1cm](b2x11) at (-0.4,12.5) [] {};
\node[fill=black, circle, draw=black, inner sep=0, minimum size=0.1cm](b2x12) at (0,12.5) [] {};
\node[fill=black, circle, draw=black, inner sep=0, minimum size=0.1cm](b2x13) at (-0.2,12.9) [] {};
\node[ circle, draw=orange, thick, inner sep=0, minimum size=0.52cm](b1c1) at (-0.2,12.7) [label=left:$C^1_{x_1^b}$] {};
\path
(b2x11) edge (b2x12)
(b2x12) edge (b2x13)
(b2x13) edge (b2x11)
(bx1) edge[color=orange,thick] (b1c1)
(bx1) edge[color=orange,thick] (b1c2);

\node[fill=black, circle, draw=black, inner sep=0, minimum size=0.1cm](bx2) at (3,11.6) [label=right:$x_{2}^{b}$] {};
\node[fill=black, circle, draw=black, inner sep=0, minimum size=0.1cm](b1x21) at (6,11.5) [] {};
\node[fill=black, circle, draw=black, inner sep=0, minimum size=0.1cm](b1x22) at (6.4,11.5) [] {};
\node[fill=black, circle, draw=black, inner sep=0, minimum size=0.1cm](b1x23) at (6.2,11.9) [] {};
\node[ circle, draw=orange, thick, inner sep=0, minimum size=0.52cm](b2c2) at (6.2,11.7) [label=right:$C^2_{x_2^b}$] {};
\path
(b1x21) edge (b1x22)
(b1x22) edge (b1x23)
(b1x23) edge (b1x21);
\node[fill=black, circle, draw=black, inner sep=0, minimum size=0.1cm](b2x21) at (-0.4,11.5) [] {};
\node[fill=black, circle, draw=black, inner sep=0, minimum size=0.1cm](b2x22) at (0,11.5) [] {};
\node[fill=black, circle, draw=black, inner sep=0, minimum size=0.1cm](b2x23) at (-0.2,11.9) [] {};
\node[ circle, draw=orange, thick, inner sep=0, minimum size=0.52cm](b2c1) at (-0.2,11.7) [label=left:$C^1_{x_2^b}$] {};
\path
(b2x21) edge (b2x22)
(b2x22) edge (b2x23)
(b2x23) edge (b2x21)
(bx2) edge[color=orange, thick] (b2c1)
(bx2) edge[color=orange, thick] (b2c2);

\node[fill=black, circle, draw=black, inner sep=0, minimum size=0.1cm](bx7) at (3,10) [label=right:$x_{7}^{b}$] {};
\node[fill=black, circle, draw=black, inner sep=0, minimum size=0.1cm](b1x71) at (6,10) [] {};
\node[fill=black, circle, draw=black, inner sep=0, minimum size=0.1cm](b1x72) at (6.4,10) [] {};
\node[fill=black, circle, draw=black, inner sep=0, minimum size=0.1cm](b1x73) at (6.2,10.4) [] {};
\node[ circle, draw=orange, thick, inner sep=0, minimum size=0.52cm](b7c2) at (6.2,10.2) [label=right:$C^2_{x_7^b}$] {};
\path
(b1x71) edge (b1x72)
(b1x72) edge (b1x73)
(b1x73) edge (b1x71);
\node[fill=black, circle, draw=black, inner sep=0, minimum size=0.1cm](b2x71) at (-0.4,10) [] {};
\node[fill=black, circle, draw=black, inner sep=0, minimum size=0.1cm](b2x72) at (0,10) [] {};
\node[fill=black, circle, draw=black, inner sep=0, minimum size=0.1cm](b2x73) at (-0.2,10.4) [] {};
\node[ circle, draw=orange, thick, inner sep=0, minimum size=0.52cm](b7c1) at (-0.2,10.2) [label=left:$C^1_{x_7^b}$] {};
\path
(b2x71) edge (b2x72)
(b2x72) edge (b2x73)
(b2x73) edge (b2x71)
(bx7) edge[color=orange, thick] (b7c1)
(bx7) edge[color=orange, thick] (b7c2);

\node[fill=black, circle, draw=black, inner sep=0, minimum size=0.1cm](by1) at (3,7.8) [label=right:$y_{1}^{b}$] {};
\node[fill=black, circle, draw=black, inner sep=0, minimum size=0.1cm](b1y11) at (6,8) [] {};
\node[fill=black, circle, draw=black, inner sep=0, minimum size=0.1cm](b1y12) at (6.4,8) [] {};
\node[fill=black, circle, draw=black, inner sep=0, minimum size=0.1cm](b1y13) at (6.2,8.4) [] {};
\node[ circle, draw=orange, thick, inner sep=0, minimum size=0.52cm](yb1c2) at (6.2,8.2) 
[label=right:$C^2_{y_1^b}$]{};
\path
(b1y11) edge (b1y12)
(b1y12) edge (b1y13)
(b1y13) edge (b1y11);
\node[fill=black, circle, draw=black, inner sep=0, minimum size=0.1cm](b2y11) at (-0.4,8) [] {};
\node[fill=black, circle, draw=black, inner sep=0, minimum size=0.1cm](b2y12) at (0,8) [] {};
\node[fill=black, circle, draw=black, inner sep=0, minimum size=0.1cm](b2y13) at (-0.2,8.4) [] {};
\node[ circle, draw=orange, thick, inner sep=0, minimum size=0.52cm](yb1c1) at (-0.2,8.2) [label=left:$C^1_{y_1^b}$] {};
\path
(b2y11) edge (b2y12)
(b2y12) edge (b2y13)
(b2y13) edge (b2y11)
(by1) edge[color=orange, thick] (yb1c1)
(by1) edge[color=orange, thick]  (yb1c2);

\node[fill=black, circle, draw=black, inner sep=0, minimum size=0.1cm](by2) at (3,7) [label=right:$y_{2}^{b}$] {};
\node[fill=black, circle, draw=black, inner sep=0, minimum size=0.1cm](b1y21) at (6,7) [] {};
\node[fill=black, circle, draw=black, inner sep=0, minimum size=0.1cm](b1y22) at (6.4,7) [] {};
\node[fill=black, circle, draw=black, inner sep=0, minimum size=0.1cm](b1y23) at (6.2,7.4) [] {};
\node[ circle, draw=orange, thick, inner sep=0, minimum size=0.52cm](yb2c2) at (6.2,7.2) 
[label=right:$C^2_{y_2^b}$]{};
\path
(b1y21) edge (b1y22)
(b1y22) edge (b1y23)
(b1y23) edge (b1y21);
\node[fill=black, circle, draw=black, inner sep=0, minimum size=0.1cm](b2y21) at (-0.4,7) [] {};
\node[fill=black, circle, draw=black, inner sep=0, minimum size=0.1cm](b2y22) at (0,7) [] {};
\node[fill=black, circle, draw=black, inner sep=0, minimum size=0.1cm](b2y23) at (-0.2,7.4) [] {};
\node[ circle, draw=orange, thick, inner sep=0, minimum size=0.52cm](yb2c1) at (-0.2,7.2) [label=left:$C^1_{y_2^b}$] {};
\path
(b2y21) edge (b2y22)
(b2y22) edge (b2y23)
(b2y23) edge (b2y21)
(by2) edge [color=orange, thick] (yb2c1)
(by2) edge [color=orange, thick] (yb2c2);

\node[fill=black, circle, draw=black, inner sep=0, minimum size=0.1cm](by7) at (3,6) [label=right:$y_{7}^{b}$] {};
\node[fill=black, circle, draw=black, inner sep=0, minimum size=0.1cm](b1y71) at (6,5.8) [] {};
\node[fill=black, circle, draw=black, inner sep=0, minimum size=0.1cm](b1y72) at (6.4,5.8) [] {};
\node[fill=black, circle, draw=black, inner sep=0, minimum size=0.1cm](b1y73) at (6.2,6.2) [] {};
\node[ circle, draw=orange, thick, inner sep=0, minimum size=0.52cm](yb7c2) at (6.2,6) [label=right:$C^2_{y_7^b}$] {};
\path
(b1y71) edge (b1y72)
(b1y72) edge (b1y73)
(b1y73) edge (b1y71);
\node[fill=black, circle, draw=black, inner sep=0, minimum size=0.1cm](b2y71) at (-0.4,5.8) [] {};
\node[fill=black, circle, draw=black, inner sep=0, minimum size=0.1cm](b2y72) at (0,5.8) [] {};
\node[fill=black, circle, draw=black, inner sep=0, minimum size=0.1cm](b2y73) at (-0.2,6.2) [] {};
\node[ circle, draw=orange, thick, inner sep=0, minimum size=0.52cm](yb7c1) at (-0.2,6) [label=left:$C^1_{y_7^b}$] {};
\path
(b2y71) edge (b2y72)
(b2y72) edge (b2y73)
(b2y73) edge (b2y71)
(by7) edge[color=orange, thick] (yb7c1)
(by7) edge[color=orange, thick] (yb7c2);

\draw(b1)--(bx1);
\draw(b1)--(bx2);
\draw(b1)--(bx7);
\draw(b1)--(by1);
\draw(b1)--(by2);
\draw(b1)--(by7);

\draw(b2)--(bx1);
\draw(b2)--(bx2);
\draw(b2)--(bx7);
\draw(b2)--(by1);
\draw(b2)--(by2);
\draw(b2)--(by7);

\node[fill=black, circle, draw=black, inner sep=0, minimum size=0.1cm](a1) at (0.4, 18) [label=left:$\color{red} a_{1}$] {};
\node[fill=black, circle, draw=black, inner sep=0, minimum size=0.1cm](a2) at (5.6, 18) [label=right:$\color{red} a_{2}$] {};

\node[fill=black, circle, draw=black, inner sep=0, minimum size=0.1cm](ax1) at (3,21.6) [label=right:$x_{1}^{a}$] {};
\node[fill=black, circle, draw=black, inner sep=0, minimum size=0.1cm](a1x11) at (6,21.5) [] {};
\node[fill=black, circle, draw=black, inner sep=0, minimum size=0.1cm](a1x12) at (6.4,21.5) [] {};
\node[fill=black, circle, draw=black, inner sep=0, minimum size=0.1cm](a1x13) at (6.2,21.9) [] {};
\node[ circle, draw=orange, thick, inner sep=0, minimum size=0.52cm](ax1c2) at (6.2,21.7) [label=right:$C_{x^a_1}^2$] {};

\path
(a1x11) edge (a1x12)
(a1x12) edge (a1x13)
(a1x13) edge (a1x11);
\node[fill=black, circle, draw=black, inner sep=0, minimum size=0.1cm](a2x11) at (-0.4,21.5) [] {};
\node[fill=black, circle, draw=black, inner sep=0, minimum size=0.1cm](a2x12) at (0,21.5) [] {};
\node[fill=black, circle, draw=black, inner sep=0, minimum size=0.1cm](a2x13) at (-0.2,21.9) [] {};
\node[ circle, draw=orange, thick, inner sep=0, minimum size=0.52cm](ax1c1) at (-0.2,21.7) [label=left:$C_{x^a_1}^1$] {};
\path
(a2x11) edge (a2x12)
(a2x12) edge (a2x13)
(a2x13) edge (a2x11)
(ax1) edge [color=orange, thick] (ax1c1)
(ax1) edge [color=orange, thick] (ax1c2);

\node[fill=black, circle, draw=black, inner sep=0, minimum size=0.1cm](ax2) at (3,20.6) [label=right:$x_{2}^{c}$] {};
\node[fill=black, circle, draw=black, inner sep=0, minimum size=0.1cm](a1x21) at (6,20.5) [] {};
\node[fill=black, circle, draw=black, inner sep=0, minimum size=0.1cm](a1x22) at (6.4,20.5) [] {};
\node[fill=black, circle, draw=black, inner sep=0, minimum size=0.1cm](a1x23) at (6.2,20.9) [] {};
\node[ circle, draw=orange, thick, inner sep=0, minimum size=0.52cm](ax2c2) at (6.2,20.7) [label=right:$C_{x^a_2}^2$] {};
\path
(a1x21) edge (a1x22)
(a1x22) edge (a1x23)
(a1x23) edge (a1x21);
\node[fill=black, circle, draw=black, inner sep=0, minimum size=0.1cm](a2x21) at (-0.4,20.5) [] {};
\node[fill=black, circle, draw=black, inner sep=0, minimum size=0.1cm](a2x22) at (0,20.5) [] {};
\node[fill=black, circle, draw=black, inner sep=0, minimum size=0.1cm](a2x23) at (-0.2,20.9) [] {};
\node[ circle, draw=orange, thick, inner sep=0, minimum size=0.52cm](ax2c1) at (-0.2,20.7) [label=left:$C_{x^a_2}^1$] {};
\path
(a2x21) edge (a2x22)
(a2x22) edge (a2x23)
(a2x23) edge (a2x21)
(ax2) edge [color=orange, thick] (ax2c1)
(ax2) edge [color=orange, thick] (ax2c2);

\node[fill=black, circle, draw=black, inner sep=0, minimum size=0.1cm](ax7) at (3,19) [label=right:$x_{7}^{a}$] {};
\node[fill=black, circle, draw=black, inner sep=0, minimum size=0.1cm](a1x71) at (6,19) [] {};
\node[fill=black, circle, draw=black, inner sep=0, minimum size=0.1cm](a1x72) at (6.4,19) [] {};
\node[fill=black, circle, draw=black, inner sep=0, minimum size=0.1cm](a1x73) at (6.2,19.4) [] {};
\node[ circle, draw=orange, thick, inner sep=0, minimum size=0.52cm](ax7c2) at (6.2,19.2) [label=right:$C_{x^a_7}^2$] {};
\path
(a1x71) edge (a1x72)
(a1x72) edge (a1x73)
(a1x73) edge (a1x71);
\node[fill=black, circle, draw=black, inner sep=0, minimum size=0.1cm](a2x71) at (-0.4,19) [] {};
\node[fill=black, circle, draw=black, inner sep=0, minimum size=0.1cm](a2x72) at (0,19) [] {};
\node[fill=black, circle, draw=black, inner sep=0, minimum size=0.1cm](a2x73) at (-0.2,19.4) [] {};
\node[ circle, draw=orange, thick, inner sep=0, minimum size=0.52cm](ax7c1) at (-0.2,19.2) [label=left:$C_{x^a_7}^1$] {};
\path
(a2x71) edge (a2x72)
(a2x72) edge (a2x73)
(a2x73) edge (a2x71)
(ax7) edge [color=orange, thick] (ax7c1)
(ax7) edge [color=orange, thick] (ax7c2);

\node[fill=black, circle, draw=black, inner sep=0, minimum size=0.1cm](ay1) at (3,17) [label=right:$y_{1}^{a}$] {};
\node[fill=black, circle, draw=black, inner sep=0, minimum size=0.1cm](a1y11) at (6,17) [] {};
\node[fill=black, circle, draw=black, inner sep=0, minimum size=0.1cm](a1y12) at (6.4,17) [] {};
\node[fill=black, circle, draw=black, inner sep=0, minimum size=0.1cm](a1y13) at (6.2,17.4) [] {};
\node[ circle, draw=orange, thick, inner sep=0, minimum size=0.52cm](ay1c2) at (6.2,17.2) [label=right:$C_{y^a_1}^2$] {};
\path
(a1y11) edge (a1y12)
(a1y12) edge (a1y13)
(a1y13) edge (a1y11);
\node[fill=black, circle, draw=black, inner sep=0, minimum size=0.1cm](a2y11) at (-0.4,17) [] {};
\node[fill=black, circle, draw=black, inner sep=0, minimum size=0.1cm](a2y12) at (0,17) [] {};
\node[fill=black, circle, draw=black, inner sep=0, minimum size=0.1cm](a2y13) at (-0.2,17.4) [] {};
\node[ circle, draw=orange, thick, inner sep=0, minimum size=0.52cm](ay1c1) at (-0.2,17.2) [label=left:$C_{y^a_1}^1$] {};
\path
(a2y11) edge (a2y12)
(a2y12) edge (a2y13)
(a2y13) edge (a2y11)
(ay1) edge [color=orange, thick] (ay1c1)
(ay1) edge [color=orange, thick] (ay1c2);

\node[fill=black, circle, draw=black, inner sep=0, minimum size=0.1cm](ay2) at (3,16) [label=right:$y_{2}^{a}$] {};
\node[fill=black, circle, draw=black, inner sep=0, minimum size=0.1cm](a1y21) at (6,16) [] {};
\node[fill=black, circle, draw=black, inner sep=0, minimum size=0.1cm](a1y22) at (6.4,16) [] {};
\node[fill=black, circle, draw=black, inner sep=0, minimum size=0.1cm](a1y23) at (6.2,16.4) [] {};
\node[ circle, draw=orange, thick, inner sep=0, minimum size=0.52cm](ay2c2) at (6.2,16.2) [label=right:$C_{y^a_2}^2$] {};
\path
(a1y21) edge (a1y22)
(a1y22) edge (a1y23)
(a1y23) edge (a1y21);
\node[fill=black, circle, draw=black, inner sep=0, minimum size=0.1cm](a2y21) at (-0.4,16) [] {};
\node[fill=black, circle, draw=black, inner sep=0, minimum size=0.1cm](a2y22) at (0,16) [] {};
\node[fill=black, circle, draw=black, inner sep=0, minimum size=0.1cm](a2y23) at (-0.2,16.4) [] {};
\node[ circle, draw=orange, thick, inner sep=0, minimum size=0.52cm](ay2c1) at (-0.2,16.2) [label=left:$C_{y^a_2}^1$] {};
\path
(a2y21) edge (a2y22)
(a2y22) edge (a2y23)
(a2y23) edge (a2y21)
(ay2) edge [color=orange, thick] (ay2c1)
(ay2) edge [color=orange, thick] (ay2c2);

\node[fill=black, circle, draw=black, inner sep=0, minimum size=0.1cm](ay7) at (3,15) [label=right:$y_{7}^{a}$] {};
\node[fill=black, circle, draw=black, inner sep=0, minimum size=0.1cm](a1y71) at (6,14.8) [] {};
\node[fill=black, circle, draw=black, inner sep=0, minimum size=0.1cm](a1y72) at (6.4,14.8) [] {};
\node[fill=black, circle, draw=black, inner sep=0, minimum size=0.1cm](a1y73) at (6.2,15.2) [] {};
\node[ circle, draw=orange, thick, inner sep=0, minimum size=0.52cm](ay7c2) at (6.2,15) [label=right:$C_{y^a_7}^2$] {};
\path
(a1y71) edge (a1y72)
(a1y72) edge (a1y73)
(a1y73) edge (a1y71);
\node[fill=black, circle, draw=black, inner sep=0, minimum size=0.1cm](a2y71) at (-0.4,14.8) [] {};
\node[fill=black, circle, draw=black, inner sep=0, minimum size=0.1cm](a2y72) at (0,14.8) [] {};
\node[fill=black, circle, draw=black, inner sep=0, minimum size=0.1cm](a2y73) at (-0.2,15.2) [] {};
\node[ circle, draw=orange, thick, inner sep=0, minimum size=0.52cm](ay7c1) at (-0.2,15) [label=left:$C_{y^a_7}^1$] {};
\path
(a2y71) edge (a2y72)
(a2y72) edge (a2y73)
(a2y73) edge (a2y71)
(ay7) edge [color=orange, thick] (ay7c1)
(ay7) edge [color=orange, thick] (ay7c2);

\path 
(y7c2) edge [color=green, thick] (ax7c1)
(yb7c2) edge [color=green, thick] (y7c1)
(x7c2) edge [color=green, thick] (ay7c1)
(ay7c2) edge [color=green, thick] (yb7c1)
(ax7c2) edge [color=green, thick] (b7c1)
(b7c2) edge [color=green, thick] (x7c1)
(a1) edge [color=yellow, thick] (b1)
(b1) edge [color=yellow, thick] (c1)
(a2) edge [color=yellow, thick] (b2)
(b2) edge [color=yellow, thick] (c2)
(a1) edge [color=yellow, thick] (a2)
(a1) edge [color=yellow, thick] (b2)
(a1) edge [color=yellow, thick] (c2)
(b1) edge [color=yellow, thick] (a2)
(b1) edge [color=yellow, thick] (b2)
(b1) edge [color=yellow, thick] (c2)
(c1) edge [color=yellow, thick] (a2)
(c1) edge [color=yellow, thick] (b2)
(c1) edge [color=yellow, thick] (c2)
(a1) edge [color=yellow, thick] (c1)
(a2) edge [color=yellow, thick] (c2);

\draw(a1)--(ax1);
\draw(a1)--(ax2);
\draw(a1)--(ax7);
\draw(a1)--(ay1);
\draw(a1)--(ay2);
\draw(a1)--(ay7);

\draw(a2)--(ax1);
\draw(a2)--(ax2);
\draw(a2)--(ax7);
\draw(a2)--(ay1);
\draw(a2)--(ay2);
\draw(a2)--(ay7);

\node[fill=black, circle, draw=black, inner sep=0, minimum size=0.1cm](a) at (-7,1.5) [label=above:$\color{red} { a}$] {};
\node[fill=black, circle, draw=black, inner sep=0, minimum size=0.1cm](b) at (-8.5,-1.5) [label=left:$\color{blue} {b}$] {};
\node[fill=black, circle, draw=black, inner sep=0, minimum size=0.1cm](c) at (-5.5,-1.5) [label=right:$\color{green} {c}$] {};
\node[fill=black, inner sep=0, minimum size=0] at (-7,-2) [label=below:$G$] {};
\path 
(a) edge (b)
(b) edge (c)
(c) edge (a);

\draw[thick,orange](ax1c2)..controls(7,21.2 )..(ax2c2);
\draw[thick,orange](ax1c1)..controls(-1,21.2 )..(ax2c1);
\draw[thick,orange,dashed](ax7c2)..controls(7,19.9 )..(ax2c2);
\draw[thick,orange,dashed](ax7c1)..controls(-1,19.9 )..(ax2c1);
\draw[thick,orange](ay1c1)..controls(-1,16.7)..(ay2c1); 
\draw[thick,orange](ay1c2)..controls(7,16.7)..(ay2c2); 
\draw[thick,orange,dashed](ay7c1)..controls(-1,15.6)..(ay2c1); 
\draw[thick,orange,dashed](ay7c2)..controls(7,15.6)..(ay2c2); 
\draw[thick,orange](yb1c2)..controls(7,7.7)..(yb2c2);
\draw[thick,orange](yb1c1)..controls(-1,7.7)..(yb2c1);
\draw[thick,orange,dashed](yb7c1)..controls(-1,6.6)..(yb2c1);
\draw[thick,orange,dashed](yb7c2)..controls(7,6.6)..(yb2c2);
\draw[thick,orange](b1c2)..controls(7,12.2)..(b2c2);
\draw[thick,orange](b1c1)..controls(-1,12.2)..(b2c1);
\draw[thick,orange,dashed](b7c1)..controls(-1,10.9)..(b2c1);
\draw[thick,orange,dashed](b7c2)..controls(7,10.9)..(b2c2);

\draw[thick,orange](y1c1)..controls(-1,-1.3)..(y2c1);
\draw[thick,orange](y1c2)..controls(7,-1.3)..(y2c2);
\draw[thick,orange,dashed](y7c2)..controls(7,-2.4)..(y2c2);
\draw[thick,orange,dashed](y7c1)..controls(-1,-2.4)..(y2c1);

\draw[thick,orange](x1c2)..controls(7,3.2)..(x2c2);
\draw[thick,orange](x1c1)..controls(-1,3.2)..(x2c1);
\draw[thick,orange,dashed](x7c1)..controls(-1,1.9)..(x2c1);
\draw[thick,orange,dashed](x7c2)..controls(7,1.9)..(x2c2);

\end{tikzpicture}
     \caption{ The reduction of
an instance $G$ of the {\sc Dominating Set} problem on circle graphs to an instance
$G'$ of the {\sc Defensive Alliance$^{\mbox{F}}$} problem in Theorem \ref{circleTheorem}. Here $2n+1=7$. One degree forbidden vertices introduced in Step 4 are not shown here. }
     \label{circleNP}
 \end{figure}
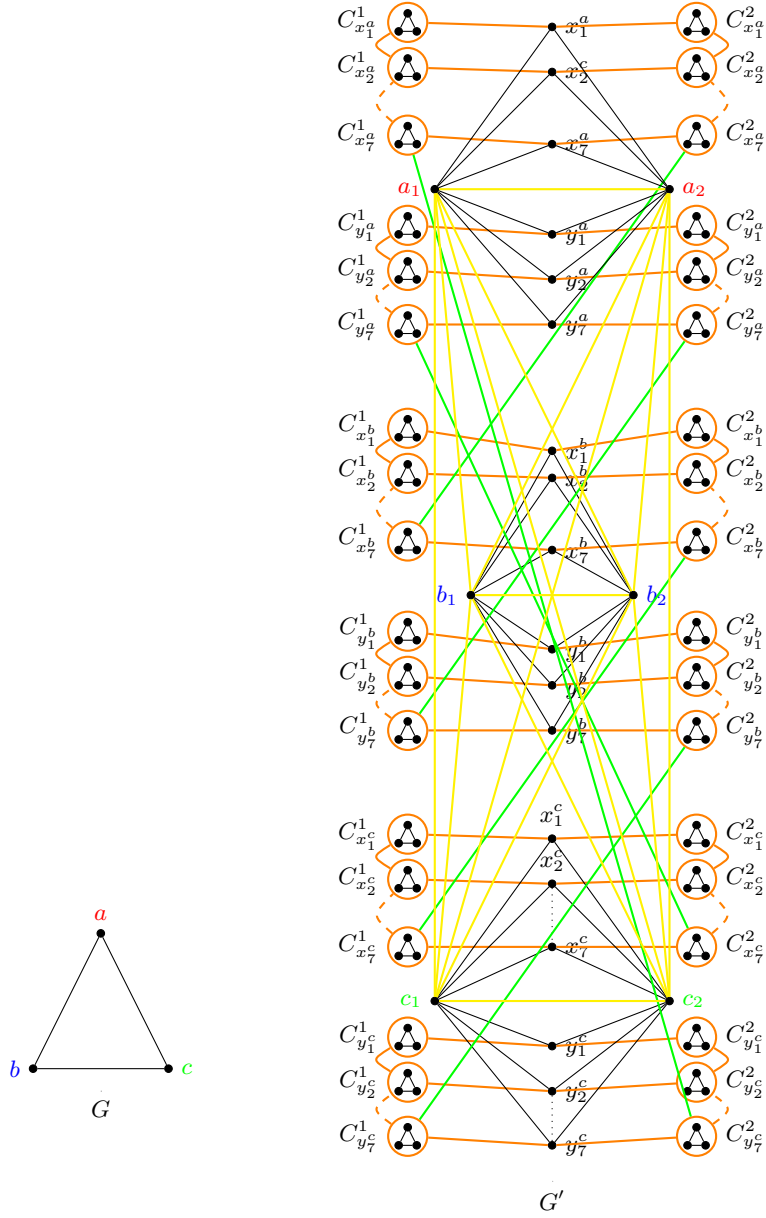

 \begin{figure}
     \centering
    \begin{tikzpicture}[scale=0.5] 
\node[fill=black, circle, draw=black, inner sep=0, minimum size=0.1cm](ay7) at (3,15) [label=above:$y_{2n+1}^{a}$] {};
\node[fill=red, circle, draw=black, inner sep=0, minimum size=0.1cm](a1y71) at (6,14.8) [] {};
\node[fill=red, circle, draw=black, inner sep=0, minimum size=0.1cm](a1y72) at (6.4,14.8) [] {};
\node[fill=red, circle, draw=black, inner sep=0, minimum size=0.1cm](a1y73) at (6.2,15.2) [] {};
\node[ circle, draw=orange, thick, inner sep=0, minimum size=0.52cm](ay7c2) at (6.2,15) [label=right:$C_{y^a_{2n+1}}^2$] {};

\node[fill=black, circle, draw=black, inner sep=0, minimum size=0.1cm](a2y71) at (-0.4,14.8) [] {};
\node[fill=black, circle, draw=black, inner sep=0, minimum size=0.1cm](a2y72) at (0,14.8) [] {};
\node[fill=black, circle, draw=black, inner sep=0, minimum size=0.1cm](a2y73) at (-0.2,15.2) [] {};
\node[ circle, draw=orange, thick, inner sep=0, minimum size=0.52cm](ay7c1) at (-0.2,15) [label=left:$C_{y^a_{2n+1}}^1$] {};
\node[fill=black, circle, draw=black, inner sep=0, minimum size=0.1cm](by7) at (3,10) [label=below:$y_{2n+1}^{b}$] {};
\node[fill=black, circle, draw=black, inner sep=0, minimum size=0.1cm](b1y71) at (6,9.8) [] {};
\node[fill=black, circle, draw=black, inner sep=0, minimum size=0.1cm](b1y72) at (6.4,9.8) [] {};
\node[fill=black, circle, draw=black, inner sep=0, minimum size=0.1cm](b1y73) at (6.2,10.2) [] {};
\node[ circle, draw=orange, thick, inner sep=0, minimum size=0.52cm](yb7c2) at (6.2,10) [label=right:$C^2_{y_{2n+1}^b}$] {};
\node[fill=azure, circle, draw=black, inner sep=0, minimum size=0.1cm](b2y71) at (-0.4,9.8) [] {};
\node[fill=azure, circle, draw=black, inner sep=0, minimum size=0.1cm](b2y72) at (0,9.8) [] {};
\node[fill=azure, circle, draw=black, inner sep=0, minimum size=0.1cm](b2y73) at (-0.2,10.2) [] {};
\node[ circle, draw=orange, thick, inner sep=0, minimum size=0.52cm](yb7c1) at (-0.2,10) [label=left:$C^1_{y_{2n+1}^b}$] {};
\path
(b2y71) edge (b2y72)
(b2y72) edge (b2y73)
(b2y73) edge (b2y71)
(by7) edge[color=orange, thick] (yb7c1)
(by7) edge[color=orange, thick] (yb7c2);
\path
(a2y71) edge (a2y72)
(a2y72) edge (a2y73)
(a2y73) edge (a2y71)
(ay7) edge [color=orange, thick] (ay7c1)
(ay7) edge [color=orange, thick] (ay7c2)
(yb7c1) edge [color=green, thick] (ay7c2);

\path
(b1y71) edge (b1y72)
(b1y72) edge (b1y73)
(b1y73) edge (b1y71);

\path
(a1y71) edge (a1y72)
(a1y72) edge (a1y73)
(a1y73) edge (a1y71);

\draw[thick] (14,12.5) circle [radius=3.5]; 

\node[fill=black, circle, draw=black, inner sep=0, minimum size=0.1cm](z1) at (17.2,14) [label=right:$y^{b}_{2n+1}$] {};

\node[fill=black, circle, draw=black, inner sep=0, minimum size=0.1cm](z2) at (14,16) [label=above:$y^{a}_{2n+1}$] {};

\node[fill=azure, circle, draw=black, inner sep=0, minimum size=0.1cm](y1) at (14.6,15.95) [label=above:] {};

\node[fill=azure, circle, draw=black, inner sep=0, minimum size=0.1cm](y2) at (15.1,15.85) [label=above:] {};

\node[fill=azure, circle, draw=black, inner sep=0, minimum size=0.1cm](y3) at (15.6,15.65) [label=above:] {};

\node[fill=red, circle, draw=black, inner sep=0, minimum size=0.1cm](y4) at (16.2,15.2) [label=above:] {};

\node[fill=red, circle, draw=black, inner sep=0, minimum size=0.1cm](y5) at (16.5,15) [label=above:] {};

\node[fill=red, circle, draw=black, inner sep=0, minimum size=0.1cm](y6) at (16.8,14.6) [label=above:] {};

\node[fill=azure, circle, draw=black, inner sep=0, minimum size=0.1cm](y7) at (17.5,12.5) [label=above:] {};

\node[fill=azure, circle, draw=black, inner sep=0, minimum size=0.1cm](y8) at (17.5,12) [label=right:$\color{azure}{C^{1}_{y^{b}_{2n+1}}}$] {};

\node[fill=azure, circle, draw=black, inner sep=0, minimum size=0.1cm](y9) at (17.4,11.5) [label=above:] {};

\node[fill=red, circle, draw=black, inner sep=0, minimum size=0.1cm](y10) at (12.6,15.7) [label=above:] {};

\node[fill=red, circle, draw=black, inner sep=0, minimum size=0.1cm](y11) at (12.2,15.5) [label=above left:$\color{red}{C^{2}_{y^{a}_{2n+1}}}$] {};

\node[fill=red, circle, draw=black, inner sep=0, minimum size=0.1cm](y12) at (11.8,15.25) [label=above:] {};

\draw[azure](y1)--(y7);
\draw[azure](y2)--(y8);
\draw[azure](y3)--(y9);
\draw[red](y4)--(y12);
\draw[red](y5)--(y11);
\draw[red](y6)--(y10);

\node[fill=black, circle, draw=black, inner sep=0, minimum size=0.1cm](z3) at (10.85,11) [label=above:] {};

\node[fill=black, circle, draw=black, inner sep=0, minimum size=0.1cm](z4) at (15.55,9.4) [label=above:] {};

\draw(z1)--(z4);
\draw(z2)--(z3);

\node[fill=black, circle, draw=black, inner sep=0, minimum size=0.1cm](y13) at (11.15,10.5) [label=above:] {};

\node[fill=black, circle, draw=black, inner sep=0, minimum size=0.1cm](y14) at (11.35,10.2) [label=left:] {};

\node[fill=black, circle, draw=black, inner sep=0, minimum size=0.1cm](y15) at (11.65,9.855) [label=above:] {};

\node[fill=black, circle, draw=black, inner sep=0, minimum size=0.1cm](y16) at (10.45,13) [label=above:] {};

\node[fill=black, circle, draw=black, inner sep=0, minimum size=0.1cm](y17) at (10.5,12.5) [label=left:$C^{1}_{y^{a}_{2n+1}}$] {};

\node[fill=black, circle, draw=black, inner sep=0, minimum size=0.1cm](y18) at (10.55,11.855) [label=above:] {};

\draw(y16)--(y13);
\draw(y17)--(y14);
\draw(y18)--(y15);

\node[fill=black, circle, draw=black, inner sep=0, minimum size=0.1cm](y19) at (16.5,10) [label=above:] {};

\node[fill=black, circle, draw=black, inner sep=0, minimum size=0.1cm](y20) at (16.9,10.45) [label=above:] {};

\node[fill=black, circle, draw=black, inner sep=0, minimum size=0.1cm](y21) at (16,9.65) [label=above:] {};

\node[fill=black, circle, draw=black, inner sep=0, minimum size=0.1cm](y22) at (14.5,9) [label=above:] {};

\node[fill=black, circle, draw=black, inner sep=0, minimum size=0.1cm](y23) at (14.9,9.1) [label=below:$C^{2}_{y^{b}_{2n+1}}$] {};

\node[fill=black, circle, draw=black, inner sep=0, minimum size=0.1cm](y24) at (14,8.95) [label=above:] {};

\draw(y22)--(y19);
\draw(y23)--(y20);
\draw(y24)--(y21);

\end{tikzpicture}
     \caption{ Illustration of Step 3.}
     \label{step3}
 \end{figure}
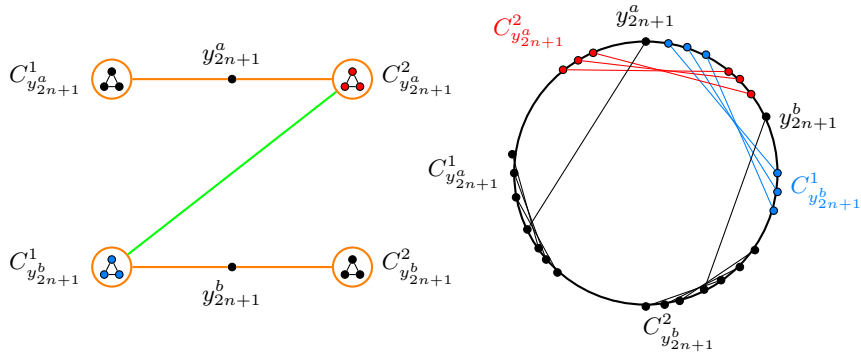

\item{\bf Step 4:} For every vertex $u$ in cliques, add $d$  forbidden vertices 
 where $d$ is the degree of $u$ until now in $G'$. For every vertex $u\in X^v\cup Y^v$, add
 six forbidden vertices and make them adjacent with  $u$. For very vertex $u\in V(G_1)\cup V(G_2)$, add $4n+3$ forbidden vertices and make them adjacent to $u$. This completes the construction of $G'$. We set $k'= 7n(4n+2)+n+k$ and $V_{\square}$ be the set of all one degree forbidden  vertices. 
\end{itemize}
\noindent We observed that the constructed graph $G'$ is indeed a circle graph, and the construction can be performed in time polynomial in $n$. We now claim that $G$ admits a dominating set 
of size at most $k$ if and only if $G'$ admits a defensive alliance $D$ of size at most $k'$ such that $D\cap V_{\square}=\emptyset$. 
Assume first that $G$ admits a dominating set $S$ of size at most $k$. 
Consider $$D = \Big\{v_1~:~v\in S\Big\} \bigcup V(G_2) \bigcup\limits_{v\in V(G)} X^v\cup Y^v \bigcup\limits_{v\in V(G)} \bigcup\limits_{i=1}^{2n+1} V(C^1_{x^v_i}) \cup V(C^2_{x^v_i}) \cup V(C^1_{y^v_i}) \cup V(C^2_{y^v_i}). $$ 
Clearly, $|D|\leq 7n(4n+2)+n+k$ and $D\cap V_{\square}=\emptyset$,  so 
it suffices to prove that $D$ is a defensive alliance in $G'$. 
We observe that every vertex in $\bigcup\limits_{v\in V(G)}\bigcup\limits_{i=1}^{2n+1} V(C^1_{x^v_i}) \cup V(C^2_{x^v_i})\cup V(C^1_{y^v_i}) \cup V(C^2_{y^v_i})$, has equal neighbours inside and outside the solution. Every  vertex 
$ v\in \bigcup\limits_{v\in V(G)} X^v\cup Y^v$ is protected as it has at least 7 neighbours inside the solution and at most 7 neighbours outside the solution. 
Each $v \in \Big\{v_1~:~v\in S\Big\} \bigcup V(G_2)$ has at least $  d+1+4n+2 $
neighbours inside the solution and at most $d+1+4n+2$ neighbours outside the solution where $d=d_{G}(x)$.  This shows that $D$ is a defensive alliance of size at most $k'$ in $G'$.

\noindent Conversely, suppose that $G'$ admits  a defensive alliance $D$ of size at most $k'$
such that $D\cap V_{\square}=\emptyset$.  We define $$V_{\triangle} =  \bigcup\limits_{v\in V(G)} X^v\cup Y^v  \bigcup\limits_{v\in V(G)} \bigcup\limits_{i=1}^{2n+1} V(C^1_{x^v_i}) \cup V(C^2_{x^v_i})\cup V(C^1_{y^v_i}) \cup V(C^2_{y^v_i}). $$ We first show that $V_{\triangle} \subseteq D$. Since $D$ in non-empty,  it should contain a vertex from either $V_{\triangle}$ or $V(G_1)\cup V(G_2)$.  We consider the following  cases: \\\\

\noindent {\it Case 1:} Suppose $D$ contains a vertex  from 
$\bigcup\limits_{i=1}^{2n+1} V(C^1_{x^v_i}) \cup V(C^2_{x^v_i})\cup V(C^1_{y^v_i}) \cup V(C^2_{y^v_i})$. Without loss of generality, we may assume that $D$ contains a vertex $u$ from $V(C^1_{x^v})$. It is easy to see that $u$ is protected if and only if  all its non-forbidden neighbours are inside $D$ because the number of forbidden neighbours of $u$ is equal to the number of non-forbidden neighbours. This implies that $x^v\in D$. 
It is easy to note that either $C_{x^v} \subseteq D$ or $C_{x^v} \cap D = \emptyset$. 
Since $d_{G'}(x^v)= 15$ and $x^v$ has $6$ forbidden neighbours, the above observation implies that $C^1_{x^v}\cup C^2_{x^v} \subseteq D$. This implies that $\bigcup\limits_{i=1}^{2n+1} C^1_{x_i^v}\cup C^2_{x_i^v} \subseteq D$. This in turn implies that $ X^v \subseteq D$. Note that $C^2_{x_{2n+1}^v}$ is  adjacent to $C^1_{x_{2n+1}^w}$  (resp. $C^1_{y_{2n+1}^w}$) for some $w\in G$ such that $u,w$ are consecutive elements in the sequence $S$ and $w$ appears for the first (resp. second) time in the sequence. Therefore $\bigcup\limits_{i=1}^{2n+1} C^1_{x_i^w}\cup C^2_{x_i^w} \subseteq D$ and also  $X^w\subseteq D$ if $w$ appears for the first time in the sequence; whereas
$\bigcup\limits_{i=1}^{2n+1} C^1_{y_i^w}\cup C^2_{y_i^w} \subseteq D$ and also  $Y^w\subseteq D$ if $w$ appears for the second time in the sequence.
Repeatedly applying the above argument, we get $V_{\triangle} \subseteq D$. \\\\

\noindent{\it Case 2:} Suppose $D$ contains a vertex  from  $\bigcup\limits_{v\in V(G)} X^v\cup Y^v$. Without loss of generality, we may assume that $D$ contains  $x^v$ from $X^v$.  We observe that the protection of $x^v$ clearly requires at least one vertex from the set $C^1_{x^v}\cup C^2_{x^v}$.  Now, Case 1 implies that $V_{\triangle} \subseteq D$. \\\\

\noindent {\it Case 3:} Suppose $D$ contains a vertex $v$ from $V(G_1)\cup V(G_2)$. The protection of $v$ requires at least one vertex from the set  $X^v\cup Y^v$. 
Now, Case 2 implies that $V_{\triangle} \subseteq D$. \\\\
Observe that $|V_{\triangle}|=7n(4n+2)$. Therefore $|D \cap (V(G_1)\cup V(G_2))|\leq n+k$. For each $v\in V(G)$, the protection of  every vertex in $ X^v\cup Y^v$ requires either $v_1$ or $v_2$ inside the solution. Since, $v_1$ and $v_2$ are twins, we can assume that $V(G_2)\subseteq D$. Let $v_2\in V(G_2)$. We see that $v_2$ has $d+1+4n+2$ neighbours (including
itself) inside the solution. The vertex $v_2$ has $4n+3$ forbidden neighbours. The only unsettled neighbours of $v_2$ are in $V(G_1)$ and $v_2$ has $d+1$ neighbours in $V(G_1)$. For protection of each $v_2\in V(G_2)$, we require at least one neighbour from $V(G_1)$ inside the solution. We can add at most $k$ vertices from $V(G_1)$ to the solution
as we have already added $7n(4n+2)+n$ vertices. Clearly, $S=V(G_1) \cap D$ is a dominating set of size at most $k$.  \qed\\

\subsection{Proof of Theorem \ref{circleTheorem}}
It is easy to see  that the problem is in NP. 
To show that the problem is NP-hard we give a polynomial reduction from 
{\sc Defensive Alliance$^{\mbox{F}}$}. Let $(G,k,V_{\square})$ be an instance of {\sc Defensive Alliance$^{\mbox{F}}$}, where $G$ is a circle graph. We construct an instance $(G',k')$ of {\sc Defensive Alliance} the following way. For every $x\in V_{\square}$, create a vertex $x'$ and 
a set of $2k'$ vertices $V^{x}_{\square}$. Make both $x$ and $x'$ adjacent to 
every vertex in $V^{x}_{\square}$. This completes the construction of $G'$. Set $k'=k$.

 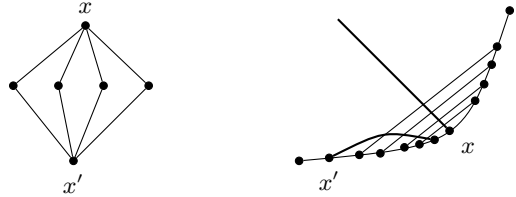
\begin{figure}
     \centering
 \begin{tikzpicture}[scale=0.4]

\node[fill=black, circle, draw=black, inner sep=0,minimum size=0.1cm](a) at (0,-5) [label=below:] {};

\node[fill=black, circle, draw=black, inner sep=0,minimum size=0.1cm](b) at (7,0) [label=below:] {};

\draw(a)..controls(5.5,-4.5)..(b);

\node[fill=black, circle, draw=black, inner sep=0,minimum size=0.1cm](x) at (5,-4) [label=below right:$x$] {};

\node(x1) at (1,0) [] {};

\draw[thick](x)--(x1);

\node[fill=black, circle, draw=black, inner sep=0,minimum size=0.1cm](z1) at (5.85,-3) [] {};

\node[fill=black, circle, draw=black, inner sep=0,minimum size=0.1cm](z2) at (4,-4.45) [] {};

\node[fill=black, circle, draw=black, inner sep=0,minimum size=0.1cm](z3) at (6.15,-2.45) [] {};

\node[fill=black, circle, draw=black, inner sep=0,minimum size=0.1cm](z4) at (3.5,-4.55) [] {};

\node[fill=black, circle, draw=black, inner sep=0,minimum size=0.1cm](z5) at (6.4,-1.8) [] {};

\node[fill=black, circle, draw=black, inner sep=0,minimum size=0.1cm](z6) at (2.7,-4.75) [] {};

\node[fill=black, circle, draw=black, inner sep=0,minimum size=0.1cm](z7) at (6.58,-1.2) [] {};

\node[fill=black, circle, draw=black, inner sep=0,minimum size=0.1cm](z8) at (2,-4.8) [] {};

\node[fill=black, circle, draw=black, inner sep=0,minimum size=0.1cm](c1) at (1,-4.9) [label=below:$x'$] {};

\node[fill=black, circle, draw=black, inner sep=0,minimum size=0.1cm](c2) at (4.5,-4.3) [] {};

\draw(z1)--(z2);
\draw(z3)--(z4);
\draw(z5)--(z6);
\draw(z7)--(z8);

\draw[thick](c1)..controls(2.8,-4)..(c2);

\node[fill=black, circle, draw=black, inner sep=0,minimum size=0.1cm](y) at (-7.1,-0.5) [label=above:$x$] {};

\node[fill=black, circle, draw=black, inner sep=0,minimum size=0.1cm](y1) at (-9.5,-2.5) [] {};
\node[fill=black, circle, draw=black, inner sep=0,minimum size=0.1cm](y2) at (-8,-2.5) [] {};
\node[fill=black, circle, draw=black, inner sep=0,minimum size=0.1cm](y3) at (-6.5,-2.5) [] {};
\node[fill=black, circle, draw=black, inner sep=0,minimum size=0.1cm](y4) at (-5,-2.5) [] {};

\node[fill=black, circle, draw=black, inner sep=0,minimum size=0.1cm](y0) at (-7.5,-5) [label=below:$x'$] {};

\draw(y)--(y1);
\draw(y)--(y2);
\draw(y)--(y3);
\draw(y)--(y4);
\draw(y0)--(y1);
\draw(y0)--(y2);
\draw(y0)--(y3);
\draw(y0)--(y4);

\end{tikzpicture}
\caption{ The circle representation to get rid of forbidden vertices when $k'=2$.}
     \label{fig:forbiddenvertex}
 \end{figure}
 We observe in Figure \ref{fig:forbiddenvertex} that the constructed graph $G'$ is indeed a circle graph, and the construction can be performed in time polynomial in $n$. We now claim that $G$ admits a 
 defensive alliance $D$ of size at most $k$ such that $D\cap V_{\square}=\emptyset$
if and only if $G'$ admits a defensive alliance $D'$ of size at most $k'$.
Assume first that $D$ is a defensive alliance of size at most $k$ in $G$ 
such that $D\cap V_{\square}=\emptyset$. Consider $D'=D$. Clearly, $D'$ 
is a defensive alliance 
of size at most $k'$ in $G'$. Conversely, suppose that $G'$ admits a  defensive alliance
$D'$ of size at most $k'$. Observe that $D' \cap \bigcup\limits_{x\in V_{\square}} V^{x}_{\square}\cup\{x,x'\}=\emptyset$. As $x$ and $x'$  are of degree $2k'$, they 
cannot be part of a defensive alliance of size at most $k'$. As $x$ and $x'$  are outside 
$D'$, the vertices in $V^{x}_{\square}$ cannot be in $D'$. Consider $D=D'$. 
Clearly, $D$ is a defensive alliance of size at most $k$ in $G$ such that $D\cap V_{\square}=\emptyset$.\qed\\


\section{Conclusions}
In this work we proved that the {\sc Defensive Alliance} problem  is W[1]-hard parameterized by 
a wide range of fairly restrictive structural parameters such as the feedback vertex set number, pathwidth,  treewidth,  treedepth, and clique width of the input graph, even when 
restricted to bipartite graph. We also proved that the problem  parameterized by the vertex cover number of the input graph does not admit a polynomial compression unless coNP $\subseteq$ NP/poly; it
 cannot be solved in time $2^{o(n)}$, unless ETH fails, and   the {\sc Defensive Alliance} problem on circle graphs is NP-complete. 
By the construction of our proofs in Section \ref{hardnesssection}, it is clear that hardness also holds for problem variants that ask for defensive alliances exactly of a given size. In the future it may be interesting to study if our ideas can be useful for different kinds of alliances from the literature such as offensive and powerful alliances.
  The parameterized complexity of  offensive and defensive alliance problems 
remain unsettled  when parameterized by other important 
structural graph parameters like twin cover and modular-width.

\bibliographystyle{abbrv}
\bibliography{bibliography}

\begin{thebibliography}{10}

\bibitem{BLIEM2018334}
B.~Bliem and S.~Woltran.
\newblock Defensive alliances in graphs of bounded treewidth.
\newblock {\em Discrete Applied Mathematics}, 251:334 -- 339, 2018.

\bibitem{CHANG2012479}
C.-W. Chang, M.-L. Chia, C.-J. Hsu, D.~Kuo, L.-L. Lai, and F.-H. Wang.
\newblock Global defensive alliances of trees and cartesian product of paths
  and cycles.
\newblock {\em Discrete Applied Mathematics}, 160(4):479 -- 487, 2012.

\bibitem{chel}
M.~Chellali and T.~W. Haynes.
\newblock Global alliances and independence in trees.
\newblock {\em Discuss. Math. Graph Theory}, 27(1):19--27, 2007.

\bibitem{marekcygan}
M.~Cygan, F.~V. Fomin, L.~Kowalik, D.~Lokshtanov, D.~Marx, M.~Pilipczuk,
  M.~Pilipczuk, and S.~Saurabh.
\newblock {\em Parameterized Algorithms}.
\newblock Springer, 2015.

\bibitem{Downey}
R.~G. Downey and M.~R. Fellows.
\newblock {\em Parameterized Complexity}.
\newblock Springer, 2012.

\bibitem{Enciso2009AlliancesIG}
R.~Enciso.
\newblock {\em Alliances in graphs: Parameterized algorithms and on
  partitioning series -parallel graphs}.
\newblock PhD thesis, USA, 2009.

\bibitem{Fernau}
H.~Fernau and D.~Raible.
\newblock Alliances in graphs: a complexity-theoretic study.
\newblock In {\em Proceeding Volume II of the 33rd International Conference on
  Current Trends in Theory and Practice of Computer Science}, 2007.

\bibitem{FERNAU2009177}
H.~Fernau, J.~A. Rodríguez, and J.~M. Sigarreta.
\newblock Offensive r-alliances in graphs.
\newblock {\em Discrete Applied Mathematics}, 157(1):177 -- 182, 2009.

\bibitem{fomin_lokshtanov_saurabh_zehavi_2019}
F.~V. Fomin, D.~Lokshtanov, S.~Saurabh, and M.~Zehavi.
\newblock {\em Kernelization: Theory of Parameterized Preprocessing}.
\newblock Cambridge University Press, 2019.

\bibitem{frick}
G.~Fricke, L.~Lawson, T.~Haynes, M.~Hedetniemi, and S.~Hedetniemi.
\newblock A note on defensive alliances in graphs.
\newblock {\em Bulletin of the Institute of Combinatorics and its
  Applications}, 38:37--41, 2003.

\bibitem{ICDCIT2021}
A.~Gaikwad, S.~Maity, and S.~K. Tripathi.
\newblock Parameterized complexity of defensive and offensive alliances in
  graphs.
\newblock In D.~Goswami and T.~A. Hoang, editors, {\em Distributed Computing
  and Internet Technology}, pages 175--187, Cham, 2021. Springer International
  Publishing.

\bibitem{mss}
R.~Ganian, F.~Klute, and S.~Ordyniak.
\newblock On structural parameterizations of the bounded-degree vertex deletion
  problem.
\newblock {\em Algorithmica}, 2020.

\bibitem{10.5555/1292785}
L.~H. Jamieson.
\newblock {\em Algorithms and Complexity for Alliances and Weighted Alliances
  of Various Types}.
\newblock PhD thesis, USA, 2007.

\bibitem{Lindsay}
L.~H. Jamieson, S.~T. Hedetniemi, and A.~A. McRae.
\newblock The algorithmic complexity of alliances in graphs.
\newblock {\em Journal of Combinatorial Mathematics and Combinatorial
  Computing}, 68:137--150, 2009.

\bibitem{soda1999}
D.~S. Johnson and M.~Szegedy.
\newblock What are the least tractable instances of max independent set?
\newblock In {\em Proceedings of the Tenth Annual ACM-SIAM Symposium on
  Discrete Algorithms}, SODA '99, page 927–928, USA, 1999. Society for
  Industrial and Applied Mathematics.

\bibitem{KEIL199351}
J.~Keil.
\newblock The complexity of domination problems in circle graphs.
\newblock {\em Discrete Applied Mathematics}, 42(1):51--63, 1993.

\bibitem{KIYOMI201791}
M.~Kiyomi and Y.~Otachi.
\newblock Alliances in graphs of bounded clique-width.
\newblock {\em Discrete Applied Mathematics}, 223:91 -- 97, 2017.

\bibitem{Kloks94}
T.~Kloks.
\newblock {\em Treewidth, Computations and Approximations}, volume 842 of {\em
  Lecture Notes in Computer Science}.
\newblock Springer, 1994.

\bibitem{kris}
P.~Kristiansen, M.~Hedetniemi, and S.~Hedetniemi.
\newblock Alliances in graphs.
\newblock {\em Journal of Combinatorial Mathematics and Combinatorial
  Computing}, 48:157--177, 2004.

\bibitem{Sparsity}
J.~Nesetril and P.~O. de~Mendez.
\newblock {\em Sparsity: Graphs, Structures, and Algorithms}.
\newblock Springer Publishing Company, Incorporated, 2014.

\bibitem{Neil}
N.~Robertson and P.~Seymour.
\newblock Graph minors. iii. planar tree-width.
\newblock {\em Journal of Combinatorial Theory, Series B}, 36(1):49 -- 64,
  1984.

\bibitem{ROD}
J.~Rodríguez-Velázquez and J.~Sigarreta.
\newblock Global offensive alliances in graphs.
\newblock {\em Electronic Notes in Discrete Mathematics}, 25:157 -- 164, 2006.

\bibitem{SIGARRETA20091687}
J.~Sigarreta, S.~Bermudo, and H.~Fernau.
\newblock On the complement graph and defensive k-alliances.
\newblock {\em Discrete Applied Mathematics}, 157(8):1687 -- 1695, 2009.

\bibitem{SIGARRETA20061345}
J.~Sigarreta and J.~Rodríguez.
\newblock On defensive alliances and line graphs.
\newblock {\em Applied Mathematics Letters}, 19(12):1345 -- 1350, 2006.

\bibitem{SIGA}
J.~Sigarreta and J.~Rodríguez.
\newblock On the global offensive alliance number of a graph.
\newblock {\em Discrete Applied Mathematics}, 157(2):219 -- 226, 2009.

\bibitem{west}
D.~B. West.
\newblock {\em Introduction to Graph Theory}.
\newblock Prentice Hall, 2000.

\end{thebibliography}

\end{document}